\documentclass[letterpaper]{JHEP3} % 10pt is ignored!

 \usepackage[T1]{fontenc}
 \usepackage[latin9]{inputenc}
\usepackage{color}

\usepackage{array}
\usepackage{setspace}
\usepackage[nosort]{cite}
 \makeatletter

\usepackage{epsfig,multicol,bbm}

\usepackage{amsmath, amssymb, graphics}

\newcommand {\mathsym}[1]{{}}

\usepackage{mathrsfs}
\usepackage{amsfonts}
\usepackage{amsmath}
\usepackage{amssymb}
\usepackage{feynmf}
\usepackage{simplewick}
\usepackage{graphicx}
\usepackage{float}
\usepackage{fancyhdr}
\usepackage{slashed}

\def\be{\begin{equation}}
\def\ee{\end{equation}}
\def\bea{\begin{eqnarray}}
\def\eea{\end{eqnarray}}

\def\nn{\nonumber}
\def\cal{\mathcal}
\def\ph{\phantom}

\global \long \def \f{\phi}

\global \long \def \p{\psi}
 
\global \long \def \pt{\tilde{\p}}

\global \long \def \l{\lambda}

\global \long \def \QQ{\mathcal{Q}}

\global \long \def \Im{\mathcal{I}}
\global \long \def \Jm{\mathcal{J}}

\newcommand\fverb{\setbox\fverbbox=\hbox\bgroup\verb}
\newcommand\fverbdo{\egroup\medskip\noindent%
			\fbox{\unhbox\fverbbox}\ }
\newcommand\fverbit{\egroup\item[\fbox{\unhbox\fverbbox}]}
\newbox\fverbbox

\def\bea{\begin{eqnarray}}
\def\eea{\end{eqnarray}}
\def\be{\begin{equation}}
\def\ee{\end{equation}}

\renewcommand{\[}{\begin{equation}}
\renewcommand{\]}{\end{equation}}

\makeatother

\global\long\global\long\def\f{\phi}

\global\long\global\long\def\e{\epsilon}

\global\long\global\long\def\ad{\dot{\alpha}}

\global\long\global\long\def\bd{\dot{\beta}}

\global\long\global\long\def\la{\lambda}

\global\long\global\long\def\p{\partial}

\global\long\global\long\def\II{\mathcal{I}}

\global\long\global\long\def\JJ{\mathcal{J}}

\global\long\global\long\def\NN{\mathcal{N}}

\global\long\global\long\def\IIh{\hat{\mathcal{I}}}

\global\long\global\long\def\topp#1{\check{#1}}

\global\long\def\QQ{\mathcal{Q}}

\global\long\def\AA{\mathcal{A}}

\global\long\def\BB{\mathcal{B}}

\global\long\def\CC{\mathcal{C}}

\global\long\def\DD{\mathcal{D}}

\global \long \def \a{\alpha}
\global \long \def \b{\beta}
\global \long \def \g{\gamma}

\global \long \def \ad{\dot{\alpha}}
\global \long \def \bd{\dot{\beta}}
\global \long \def \gd{\dot{\gamma}}

\global \long \def \e{\epsilon}

\global \long \def \ab{\textbf{a}}
\global \long \def \bb{\textbf{b}}
\global \long \def \cb{\textbf{c}}
\global \long \def \db{\textbf{d}}
\global \long \def \dag{\dagger}
\global \long \def \lbr{\lbrack}
\global \long \def \rbr{\rbrack}
\global \long \def \ph{\phantom}

\global \long \def \dag{\dagger}
\global \long \def \lra{\leftrightarrow}

\newcommand{\itil}{{\tilde \imath}} 
\newcommand{\jtil}{{\tilde \jmath}}

\title{The Complete One-loop Spin Chain of ${\cal N} =1$ SQCD}

\preprint{YITP-SB-11-40
}

\author{Pedro Liendo$^{a}$\footnote{Email: pedro.liendo@stonybrook.edu}$\,$
 and   Leonardo Rastelli$^{a}$\footnote{Email: leonardo.rastelli@stonybrook.edu}
\\
\\
\it $^a$ C.N. Yang Institute for Theoretical Physics,\\
\it Stony Brook University, \\
\it Stony Brook, NY 11794-3840, USA
}
\bigskip

 \abstract{
 \vspace{0.2cm}
 
We evaluate the complete planar one-loop dilation operator
of ${\cal N}=1$ Super QCD, at  the large $N$  Banks-Zaks fixed point near the upper edge of the superconformal window.
 The spin-chain Hamiltonian turns out to be entirely fixed by the constraints of superconformal symmetry, as in ${\cal N}=4$ Super Yang-Mills
 and in ${\cal N}=2$ SuperConformal QCD. 
 }
 

\global \long \def \f{\phi} 

\global \long \def \p{\psi} 
\global \long \def \pt{\tilde{\p}}

\global \long \def \l{\lambda}

\global \long \def \QQ{\mathcal{Q}}

\global \long \def \Im{\mathcal{I}}
\global \long \def \Jm{\mathcal{J}}
\global \long \def \Nm{\mathcal{N}}
\global \long \def \Fm{\mathcal{F}}
\global \long \def \Bm{\mathcal{B}}
\global \long \def \Cm{\mathcal{C}}

\global \long \def \Vm{\mathcal{V}}

\global \long \def \Dm{\mathcal{D}}
\global \long \def \Lm{\mathcal{L}}
\global \long \def \Qm{\mathcal{Q}}
\global \long \def \Km{\mathcal{K}}
\global \long \def \Sm{\mathcal{S}}
\global \long \def \Pm{\mathcal{P}}

\global \long \def \Xm{\mathcal{X}}

\global \long \def \dag{\dagger}

\begin{document} 

\maketitle

\section{Introduction}

Superconformal  field theories have come to occupy a central position in theoretical physics.
Much has been learnt, especially about the most supersymmetric models
 in four and three dimensions, but some of the most interesting questions remain unanswered.
 To mention a glaring example, which is the focus of this paper, no substantial progress has been made over the years
 on the simplest four-dimensional ${\cal N}=1$ model originally discovered by Seiberg \cite{Seiberg:1994pq}:
$SU(N_c)$ super QCD in the conformal window. 
No weakly-coupled AdS gravity dual description is expected to exist for this class of theories; non-critical string duals
(which have an intrinsically  strongly-coupled worldsheet description)
may well exist
(see {\it e.g.} \cite{Klebanov:2004ya,Fotopoulos:2005cn,Ashok:2005py,Bigazzi:2005md,Murthy:2006xt}), but these proposals are  not yet well understood. And Seiberg duality remains largely a mystery.\footnote{We should however mention the new  precision checks of Seiberg duality for the protected spectrum, obtained with the help of  the superconformal index \cite{Romelsberger:2005eg,Romelsberger:2007ec,Dolan:2008qi}.}

 A strategy that has proved very fruitful in the case of ${\cal N}=4$ super Yang-Mills theory
 is the systematic perturbative evaluation of the planar dilation operator \cite{Minahan:2002ve,Beisert:2003tq,Beisert:2003jj,Beisert:2003ys,Zwiebel:2005er,Sieg:2010tz,Zwiebel:2011bx}, which turns out to be described by an integrable spin chain (see {\it e.g.} \cite{Beisert:2003yb,Staudacher:2004tk,Beisert:2005fw,Beisert:2005tm,Beisert:2006ez, Gromov:2009tv} for
 a very partial list of references and \cite{Beisert:2010jr} for a comprehensive review). 
  One does not {\it a priori}  expect the remarkable integrability properties of ${\cal N}=4$ SYM to fully persist in less symmetric  theories,
 but one cannot really know until one tries.\footnote{There has been much work on  integrability in $4d$ gauge theories with 
 ${\cal N}<4$, see {\it e.g.} \cite{Belitsky:2004sf,Belitsky:2005bu,Belitsky:2006av} and the recent review \cite{Korchemsky:2010kj}.
 Most investigations to date have focused on theories in the standard 't Hooft limit, where the number of fundamental flavors $N_f$ is zero or anyway kept fixed as the number of colors $N_c$ is sent to infinity. The 't Hooft limit has the drawback that conformal invariance
 is inevitably broken (except of course in ${\cal N}=4$ SYM). The breaking of conformal invariance
 does not affect  the one-loop dilation operator, but it kicks in  at two loops.}
 For   ${\cal N}=1$ SQCD, 
 a perturbative expansion is meaningful in the large $N$ Veneziano limit \cite{Veneziano:1976wm}
 of $N_c \to \infty$, $N_f \to \infty$ with $N_f /N_c$ fixed,  near the upper edge of the superconformal window $N_f \lesssim 3 N_c$. If one defines
 \be
 \frac{N_f}{ N_c} = 3 - \epsilon \, ,
 \ee 
 the large $N$ theory flows for $\epsilon \ll 1$ to a weakly-coupled Banks-Zaks fixed point \cite{Banks:1981nn}, with 't~Hooft coupling $g_{YM}^2 N_c 
\sim \epsilon$. The one-loop planar dilation operator captures the spectrum of the theory at this isolated fixed point,
while higher-loop  corrections (reorganized in powers of $\epsilon$) correspond to moving down the conformal window. 
The dual ``magnetic'' theory admits a  perturbative expansion starting 
 from the lower edge of the conformal window $N_f \gtrsim \frac{3}{2}N_f$, with a Banks-Zaks fixed point that is weakly-coupled for $\tilde \epsilon \ll 1$,
 where 
 \be
  \frac{ N_f}{  N_c} = \frac{3}{2} + \tilde \epsilon \, .
  \ee
A complete large $N$ solution of SQCD  would entail determining the dilation operator
of the electric theory to all orders in $\epsilon$, and that of the magnetic theory to all orders in $\tilde \epsilon$.
The  resummations of both expansions should then coincide --  in the ultimate triumph of Seiberg duality. Needless to say, this is a tall order, and one cannot hope to fulfill
this program  unless integrability comes to the rescue.

Recently  the one-loop dilation operator of ${\cal N}=1$ SQCD in the Veneziano limit has been determined in the scalar subsector \cite{Poland:2011kg},
and shown to coincide with the Ising spin chain in a transverse magnetic field, one of the best known integrable models.
This is a tantalizing hint, well-worth subjecting to more stringent tests.
  In this paper we determine the {\it complete} planar one-loop spin chain Hamiltonian of ${\cal N}=1$ SQCD
and make a preliminary investigation of its integrability. 

We eschew direct Feynman diagram calculations and  rely instead on symmetry.
The constraints of  superconformal invariance are sufficient to completely fix the one-loop planar dilation operator
of ${\cal N}=4$ super Yang-Mills \cite{Beisert:2003jj,Beisert:2004ry}, and also, somewhat unexpectedly, of ${\cal N}=2$ superconformal QCD, as recently shown in \cite{Liendo:2011xb}.
The calculation for ${\cal N}=1$ SQCD proceeds along similar lines as ${\cal N}=2$ SCQCD,
and again we are able to fix the one-loop Hamiltonian  from symmetry considerations alone. This is a nice  surprise.
For ${\cal N}=4$ SYM at each site of the chain sits a single irreducible multiplet,
which moreover has a simple tensor product with itself; 
by contrast in the ${\cal N}=2$ and ${\cal N}=1$ cases 
each site hosts a handful of irreducible representations, and their  tensor products have a more complicated decomposition,
leading to a rather intricate mixing problem. Despite these complications, the general structure of the calculation is the same as in ${\cal N}=4$ SYM:
 the full Hamiltonian can be uplifted from the Hamiltonian in a simple subsector, which is in turn uniquely fixed by a centrally extended $SU(1|1)$ symmetry.

The rest of the paper is organized as follows.
We begin in section \ref{preliminaries}
by introducing  ${{\cal N}=1}$ SQCD and its spin chain. We discuss the decomposition of the single-letter state space  into irreducible representations of the
 ${\cal N}=1$ superconformal  algebra and give the tensor products of any two such representations.
In section \ref{Algebraic} we show how the full planar one-loop Hamiltonian can be uplifted from a closed subsector with $SU(1,1) \times U(1|1)$ symmetry,
where it is fixed (up to overall normalization) by the centrally extended superalgebra.
By this route we obtain an expression for the full Hamiltonian in terms of the superconformal projectors onto the different
irreps that appear in the two-site state space. By making contact with the scalar sector results of \cite{Poland:2011kg} we fix
the overall normalization.
Finally we obtain the ``harmonic action'' form
of the full Hamiltonian -- a completely explicit oscillator representation that can be 
easily implemented on any state. In section \ref{Tables} we present a preliminary investigation
of the integrability of the one-loop spin chains for ${\cal N}=2$ SCQCD and ${\cal N}=1$ SQCD.
We show that both models admit a ``parity'' symmetry that commutes with the Hamiltonian
to all loops. In ${\cal N}=4$ SYM, the existence of degenerate parity pairs provided early circumstantial evidence for integrability.
We diagonalize the spectrum and search  for parity pairs, both in the ${\cal N}=2$ and ${\cal N}=1$ spin chains (in the subsectors that
were used to uplift the full Hamiltonians, for states of length $L \leq 5$). 
We find that the appearance of parity pairs is much less
systematic than in ${\cal N}=4$ SYM.
We conclude in section \ref{discussion} with a brief discussion.
Two appendices contain background material on the ${\cal N}=1$ superconformal algebra and its oscillator representation,
and the explicit expressions of the two-site superconformal primaries.

\section{Preliminaries}

\label{preliminaries}

We consider ${\cal N}=1$ SQCD, the ${\cal N}=1$ supersymmetric Yang-Mills theory with gauge group $SU(N_c)$
and $N_f$ flavors of fundamental quarks. In Table \ref{fieldcontent} we recall the familiar symmetries of the theory
and set our notations.  
  Besides the ${\cal N}=1$ vector multiplet
$(A_{\alpha \dot \alpha}$, $\l_{\a})$,
 in the adjoint representation of the gauge group,
there are two sets of
 $N_f$ chiral multiplets,  $(Q, \psi_\alpha)$ and $(\tilde Q, \tilde \psi_\alpha)$,
 respectively in the fundamental and  antifundamental
 representations of $SU(N_c)$. 
  The color and flavor  structure  is then 
 \be
(A^{a}_{\; b} , \lambda^a_{\;b}) \, , \qquad (Q^{a i} , \psi^{a i} ) \, ,\qquad  (\tilde Q_{a \itil} , \tilde \psi_{a\itil} ) \, , 
 \ee
 where $a=1, \dots N_c$ are color indices, and  $i = 1, \dots N_f $ and $\tilde \imath = 1, \dots N_f$ two independent sets of flavor indices,
 corresponding to the independent flavor
 symmetries of the  gauge-fundamental and of the gauge-antifundamental chiral multiplets.

 \begin{table}[h]
 \begin{centering}
 \begin{tabular}{|c||c|c|c|c|c|c|}
 \hline 
 & $SU(N_{c})$  & $SU(N_{f})$ & $SU(N_{f})$ & $U(1)_{B}$  & $U(1)_{r}$ \tabularnewline
 \hline
 \hline 
$\Qm_{\a}$  &  \textbf{$ \mathbf{1}$}  &  \textbf{$ \mathbf{1}$} &  \textbf{$ \mathbf{1}$}  &$0$ &$-1$ \tabularnewline
 \hline 
$\Sm^{\a}$ &  \textbf{$ \mathbf{1}$}  &  \textbf{$ \mathbf{1}$}  &  \textbf{$ \mathbf{1}$}   &$0$ & $1$ \tabularnewline
 \hline
 \hline 
$ \l_{\a}$  & Adj  &  \textbf{$ \mathbf{1}$} &  \textbf{$ \mathbf{1}$} &   $0$   &   $1$ \tabularnewline
 \hline 
$A_{\alpha \dot \alpha}$  & Adj  &  \textbf{$ \mathbf{1}$} &  \textbf{$ \mathbf{1}$}  &  $0$ &  $0$ \tabularnewline
 \hline
$Q$  & $ \Box$  & $ \Box$  &  \textbf{$ \mathbf{1}$}  &  $1$  &  $1-\frac{N_c}{N_f}$ \tabularnewline
 \hline 
$\psi_{\alpha}$  & $ \Box$ & $ \Box$    &  \textbf{$ \mathbf{1}$} &  $1$ &  $-\frac{N_c}{N_f}$ \tabularnewline
 \hline 
$\tilde Q$ & $ \overline{\Box}$   &  \textbf{$ \mathbf{1}$}  & $ \overline{\Box}$  &  $-1$  &  $1-\frac{N_c}{N_f}$ \tabularnewline
 \hline 
$ \tilde{\psi}_{\alpha}$ & $ \overline{\Box}$  & \textbf{$ \mathbf{1}$}  & $ \overline{\Box}$  &  $-1$ &  $-\frac{N_c}{N_f}$ \tabularnewline
 \hline
 \end{tabular}
 \par \end{centering}
 \caption{Field content and symmetries of  $ \Nm=1$ SQCD. We use $\alpha = \pm$ and $\dot \alpha = \dot \pm$ for Lorentz
 spinor indices. ${\cal Q}_\alpha$ and ${\cal S}^\alpha$ denote respectively the Poincar\'e and conformal supercharges.
Conjugate objects such as $\bar \lambda_{\dot \alpha}$ are not written explicitly.}
\label{fieldcontent}
 \end{table}
 
 In the  large $N$ Veneziano limit, the basic flavor-singlet local gauge-invariant operators are ``generalized single-traces'' \cite{Gadde:2009dj,Gadde:2010zi}, of the schematic form
 \be \label{gensingletrace}
 {\rm Tr}\left( \phi^{k_1} {\cal M}^{k_2} \phi^{k_3} {\cal M}^{k_4} \dots \right)\,.
 \ee
Here $\phi$ denotes any of the color-adjoint ``letters'', for example
$\phi^a_{\; b} = (\Dm^n \lambda)^{a}_{\;b}$, where $\Dm$ is a gauge-covariant derivative,
while ${\cal M}^{a}_{\, b}$ is any of the gauge-adjoint composite objects obtained by the flavor contraction of a fundamental and an antifundamental
letter, for example ${\cal M}^{a}_{\; b}= Q^{a i } \bar Q_{b i}$ or ${\cal M}^{a}_{\; b}= \bar {\tilde \psi}^{a \itil } \tilde Q_{b \itil}$. In the Veneziano limit these
are the building blocks:  a generic gauge-singlet operator factorizes into products of generalized single-trace operators,
 up to $1/N$ corrections. To leading  order in the  large $N$ limit (the planar theory) generalized single-trace operators are closed under the action
 of the dilation operator, which takes the familiar form of a spin chain Hamiltonian; as usual, the locality of the Hamiltonian (nearest neighbor, next-to-nearest-neighbor, ...)
 is related to the loop order of  the planar perturbative expansion.
 
One can also consider operators with open flavor indices of the schematic forms
\be \label{4open}
\bar q_i \, \phi^{k_1} {\cal M}^{k_2} \dots q^j \, ,\qquad \bar q_i \, \phi^{k_1} {\cal M}^{k_2} \dots \bar {\tilde q}^\jtil \, ,\qquad \tilde q_\itil \, \phi^{k_1} {\cal M}^{k_2} \dots q^j \, ,\qquad\tilde q_\itil\,  \phi^{k_1} {\cal M}^{k_2} \dots \bar  {\tilde q}^\jtil \, ,
\ee
where $q$ and $\tilde q$ stand for any of the (anti)fundamental letters. It is also understood that for operators
with index structures $_i^{\ph{ab}j}$ and $_\itil^{\ph{ab}\jtil}$ we are projecting into the adjoint representation
of $SU(N_f)$ by removing the flavor-trace term. In the Veneziano limit operators of the form (\ref{4open}) are closed under renormalization
and they can be viewed as open spin chains.

\subsection{Superconformal representations}

We are going to make crucial use of superconformal symmetry to constraint the form of the one-loop dilation operator
(the spin chain Hamiltonian). 
The letters that occupy each site of the ${\cal N}=1$ SQCD  chain belong to four distinct irreducible representations of the $SU(2,2|1)$ superconformal algebra.
We denote them by $\Xm$ (chiral multiplet), $\bar{\Xm}$ (antichiral multiplet), $\Vm$ (vector multiplet) and $\bar{\Vm}$ (conjugate vector multiplet).
This is in contrast with ${\cal N}=4$ SYM, where the letters belong to a single irreducible representation,
but rather similar to ${\cal N}=2$ SCQCD, where the single-site state space decomposes in the sum of three different representations \cite{Liendo:2011xb}.  

At one loop, the Hamiltonian is of nearest-neighbor form. The Hamiltonian
density acts on two adjacent sites and can be written as a sum
of projectors onto the irreducible representations that span the two-site state space. Because of the index structure of the spin chain,
not all orderings of two single-site representations are allowed in the two-site state space.
For example, it is not possible to have two $Q$s adjacent to each other because there is no way in which to contract the indices,
so two adjacent $\Xm$ representations are not allowed.
 On the other hand, $Q$ and $\bar{Q}$ can be placed together and, in fact, there are two ways in which this can be done, we can contract either adjacent \textit{gauge} indices or adjacent \textit{flavor} indices. 
The gauge-contracted combinations are
 (the order matters):
\begin{align}
	\Vm \times  \Vm & \qquad \bar{\Vm} \times \Vm & \bar{\Vm} \times  \bar{\Vm} & \qquad \Vm \times \bar{\Vm}
	\\
	\Vm \times \Xm & \qquad \tilde{\Xm} \times \Vm & \bar{\Vm} \times  \Xm & \qquad \tilde{\Xm} \times \bar{\Vm}
	\\
	\Vm \times \bar{\tilde{\Xm}}  & \qquad \bar{\Xm} \times \Vm & \bar{\Vm} \times   \bar{\tilde{\Xm}} & \qquad \bar{\Xm} \times 
	\bar{\Vm}
	\\
	 \bar{\Xm} \times \Xm & \qquad  \bar{\Xm} \times \bar{\tilde{\Xm}}  & \tilde{\Xm} \times \Xm  & \qquad   \tilde{\Xm} \times 
	 \bar{\tilde{\Xm}}&\, ,
	\end{align}
while the flavor-contracted combinations are:
\begin{align}
& \Xm \times \bar{\Xm} &  \bar{\tilde{\Xm}} \times \tilde{\Xm}  \,.
\end{align}
For clarity we have added a ``tilde'' to distinguish the fundamental from the antifundamental chiral multiplets, though of course
this is a distinction that pertains to the color and flavor structure,  not the superconformal structure ($\Xm$ and $\tilde \Xm$ are isomorphic
as  superconformal representations).

The classification of multiplets of the ${\cal N}=1$ superconformal algebra is reviewed in Appendix A. We follow 
 the notations of \cite{Dolan:2008qi}, according to which the multiplets that span the single-site state space are given by
\be 
\Xm  =\tilde{\Xm}= \bar{\Dm}_{(0,0)}\,, \hspace{1cm} \bar{\Xm}  =\bar{\tilde{\Xm}} =\Dm_{(0,0)}, \hspace{1cm} \Vm = \bar{\Dm}_{(\frac{1}{2},0)}\,, \hspace{1cm} \bar{\Vm} = \Dm_{(0,\frac{1}{2})}\,.
\ee
Using superconformal characters it is not difficult to decompose the tensor products of any two such multiplets into irreducible representations. We find
\begin{align}
\label{XtXT}
\tilde{\Xm} \times \Xm & =  \bar{\Bm}_{\frac{4}{3}(0,0)}\oplus\bigoplus_{q=0}^{\infty} \hat{\Cm}_{(\frac{q+1}{2},\frac{q}{2})}\, ,
\\
\bar{\Xm} \times \bar{\tilde{\Xm}} & = \Bm_{-\frac{4}{3}(0,0)}\oplus\bigoplus_{q=0}^{\infty} \hat{\Cm}_{(\frac{q}{2},\frac{q+1}{2})}\, ,
\\
\Xm \times \bar{\Xm}=\bar{\Xm} \times \Xm & =  \bigoplus_{q=0}^{\infty} \hat{\Cm}_{(\frac{q}{2},\frac{q}{2})}
=\tilde{\Xm} \times \bar{\tilde{\Xm}} =  \bar{\tilde{\Xm}} \times \tilde{\Xm}\, , 
\\
\Vm \times \Xm & =  \bar{\Bm}_{\frac{5}{3}(\frac{1}{2},0)}\oplus\bigoplus_{q=1}^{\infty} \hat{\Cm}_{(\frac{q+1}{2},\frac{q-1}{2})}=\tilde{\Xm} \times \Vm\, ,
\\
\bar{\Vm} \times \Xm  & =  \bigoplus_{q=0}^{\infty} \hat{\Cm}_{(\frac{q}{2},\frac{q+1}{2})}=\tilde{\Xm} \times \bar{\Vm}\, ,
\\
\bar{\Xm} \times \Vm & =  \bigoplus_{q=0}^{\infty} \hat{\Cm}_{(\frac{q+1}{2},\frac{q}{2})}=\Vm  \times \bar{\tilde{\Xm}}\, ,
\\
\bar{\Xm} \times \bar{\Vm} & =  \Bm_{-\frac{5}{3}(0,\frac{1}{2})}\oplus\bigoplus_{q=1}^{\infty} \hat{\Cm}_{(\frac{q-1}{2},\frac{q+1}{2})}=\bar{\Vm} \times \bar{\tilde{\Xm}}\, ,
\\
\Vm \times \Vm & =  \bar{\Bm}_{2(0,0)} \oplus\bar{\Bm}_{2(1,0)} \oplus \bigoplus_{q=2}^{\infty}\hat{\Cm}_{(\frac{q+1}{2},\frac{q-2}{2})}\, ,
\\
\bar{\Vm} \times \bar{\Vm} & =   \Bm_{-2(0,0)}\oplus \Bm_{-2(0,1)} \oplus \bigoplus_{q=2}^{\infty}\hat{\Cm}_{(\frac{q-2}{2},\frac{q+1}{2})}\, ,
\\
\label{VVbT}
\Vm \times \bar{\Vm} & =   \bigoplus_{q=1}^{\infty}\hat{\Cm}_{(\frac{q}{2},\frac{q}{2})}=\bar{\Vm} \times \Vm\, .
\end{align}

\section{Algebraic Evaluation of the Hamiltonian}
\label{Algebraic}

The evaluation of the one-loop Hamiltonian proceeds  in three steps. First,
we identify a closed $SU(1,1) \times U(1|1)$ subsector and determine
the Hamiltonian in the subsector by using the constraints of the centrally-extended symmetry.
We then uplift the result to the complete theory and obtain the full Hamiltonian as a sum of superconformal
projectors. Finally we rewrite the Hamiltonian in an explicit ``harmonic action'' form.

\subsection{The $SU(1,1) \times U(1|1)$ subsector}

Consider the subsector generated by the letters
 \bea
\label{lF}
\lambda_k = \frac{\Dm^k}{k!} \lambda_{+}\,, &  \quad & \bar{\Fm}_k  = \frac{\Dm^k}{k!}  \bar{\Fm}_{\dot + \dot +}\,,
\\
\label{QPsi}
Q_k = \frac{\Dm^k}{k!} Q\, , & \quad & \bar{\psi}_k = \frac{\Dm^k}{k!}\bar{\psi}_{\dot +}\,,
\\
\label{QPsit}
\tilde{Q}_k = \frac{\Dm^k}{k!} \tilde{Q}\,, & \quad & \bar{\tilde{\psi}}_k = \frac{\Dm^k}{k!}\bar{\tilde{\psi}}_{\dot +}\,.
\eea
with $\Dm \equiv \Dm_{+ \dot +}$. By using conservation of the engineering dimension, of the Lorentz spins and of the R-charge it is easy to see
 that this sector is closed to all loops under the action of the dilation operator. Moreoever,
the one-loop Hamiltonian  restricted to this subsector can be uplifted to the full one-loop Hamiltonian,
as each of the modules appearing on the right hand side of the tensor products (\ref{XtXT}--\ref{VVbT}) contains a representative within the subsector.
The representatives are primaries of $SU(1,1)$,
and  descendants with respect to the full $SU(2,2|1)$ algebra. 

To obtain the Hamiltonian in the subsector one could perform an explicit Feynman diagram calculation.
Instead we use a purely algebraic approach that uses the restrictions imposed by superconformal symmetry. The algebraic method was successfully used in \cite{Beisert:2004ry} and \cite{Liendo:2011xb} to find the dilation operator of $\Nm=4$ SYM and $\Nm=2$ SCQCD respectively.

The subgroup of the superconformal group acting on the sector is $SU(1,1) \times U(1|1)$. The  $SU(1,1)$ generators are
	\bea \label{J+}
	\Jm'_+(g) & = & \Pm_{+ \dot{+}}(g)\, ,
	\\
	\label{J-}
	\Jm'_-(g) & = & \Km^{+\dot{+}}(g)\, ,
	\\
	\label{J3}
	\Jm'_3(g) & = & \frac{1}{2}D_0+\frac{1}{2}\delta D(g) + \frac{1}{2}\Lm^{\ph{+}+}_{+} + \frac{1}
	{2}\dot{\Lm}^{\ph{+}\dot{+}}_{\dot{+}} \, .
	\eea
The $U(1|1)$ generators are given by
	\be
	L = \Lm_{-}^{\ph{-}-}+\dot{\Lm}_{\dot{-}}^{\ph{\dot{-}}\dot{-}}+D_0\, , \quad \bar{\Qm}(g) = \bar{Q}_{\dot{-}}(g)\, , \quad \bar{\Sm}(g) = \bar{\Sm}_{\dot{-}}(g)\, , \quad \delta D(g)
	\ee
	and can be checked to commute with the $SU(1,1)$ generators (\ref{J+}--\ref{J3}).	
	Their (anti)commutators are\footnote{The bar in $\bar{\Qm}$ and $\bar{\Sm}$ does not denote complex conjugation, we are going to impose the appropriate hermiticity condition below.}
	\begin{align}
	 [L,\bar{\Qm}(g) ] & = \bar{\Qm}(g)\, , 
	 \end{align}
	 \begin{align}
	 \lbrack L,\bar{\Sm}(g) ] & =  -\bar{\Sm}(g)\, ,
	 \\ \label{QS}
	 \{\bar{\Sm}(g),\bar{\Qm}(g)\} & =  \frac{1}{2}\delta D(g)\, .
	\end{align}
	The generator $L$ can be identified with the length operator, this implies that $\bar{\Qm}$ increases the length of a chain while $\bar{\Sm}$ decreases it.
	
	\subsection{First order expressions for $\Qm(g)$ and $\Sm(g)$}
	 
	 The procedure is now very similar to the one followed in \cite{Liendo:2011xb}  for ${\cal N}=2$ SCQCD and we shall be brief.
	 Writing $\Qm(g)=g\Qm+O(g^2)$, we formulate an ansatz for the action of the supercharges on the states of the sector compatible with the quantum numbers of the fields and impose invariance  under the $SU(1,1)$ algebra, $[\Jm',\bar{\Qm}]=0$ and $[\Jm',\bar{\Sm}]=0$. In fact strict invariance is too restrictive and one needs  only to impose 
	 vanishing of these commutators up to  local gauge transformations on the chain. 
	 It can be easily checked that the following transformations evaluate to zero on any closed chain,
	\bea 
	\label{Qactfirst}
	[\Jm'_+, \bar{\Qm}] \l_n & = & \a \left( \l_n \l + \l\l_n \right)\, ,
	\\
	\lbrack \Jm'_+, \bar{\Qm}] \bar{\Fm}_n & = &\a \left( -\bar{\Fm}_n \l + \l\bar{\Fm}_n \right)\, ,
	\\
	\lbrack \Jm'_+, \bar{\Qm}] Q_n & = & \a \l Q_n \, ,
	\\
	\lbrack \Jm'_+, \bar{\Qm}] \bar{\psi}_n & = & \a \bar{\psi}_n \l \, ,
	\\
	\lbrack \Jm'_+, \bar{\Qm}] \tilde{Q}_n & = & -\a \tilde{Q}_n \l \, ,
	\\
	 \label{Qactlast}
	\lbrack \Jm'_+, \bar{\Qm}] \bar{\tilde{\psi}}_n & = & \a \l \bar{\tilde{\psi}}_n\,,
	\eea
	where $\a$ is an arbitrary gauge parameter. 
	The action of $\bar{\Qm}$ consistent with these transformations is
	\bea 
	\bar{\Qm} \l_n & = & \a \sum_{k'=0}^{n-1}\frac{n+1}{(k'+1)(n-k')}\l_{k'} \l_{n-k'-1}\, ,	
	\\
	\nn
	\bar{\Qm} \bar{\Fm}_n & = & \a \sum_{k'=0}^{n-1} \left(-\frac{1}{n-k'}\bar{\Fm}_{k'}\l_{n-k'-1} 
	+\frac{1}{k'+1}\l_{k'} \bar{\Fm}_{n-k'+1}     \right)
	\\
	& & + \a' \sum_{k'=0}^{n} Q_{k'}^i\bar{\psi}_{n-k'\,i} + \a'' \sum_{k'=0}^{n} \bar{\tilde{\psi}}_{k'}^{\tilde{\imath}}\tilde{Q}_{n-k'\, \tilde{\imath}}\, ,
	\\
	\bar{\Qm} Q_n & = & \a \sum_{k'=0}^{n-1}\frac{1}{k'+1}\l_{k'} Q_{n-k'-1}\, ,
    \\
	\bar{\Qm} \bar{\p}_n & = & \a \sum_{k'=0}^{n-1}\frac{1}{n-k'}\bar{\p}_{k'} \l_{n-k'-1}\, ,
	\\
	\bar{\Qm} \tilde{Q}_n & = & -\a \sum_{k'=0}^{n-1}\frac{1}{n-k'}\tilde{Q}_{k'}\l_{n-k'+1} \, ,
    \eea
    \bea
	\bar{\Qm} \bar{\pt}_n & = & \a \sum_{k'=0}^{n-1}\frac{1}{k'+1}\l_{k'}\bar{\p}_{n-k'+1}\, .
	\eea	
	The terms with $\a'$ and $\a''$ are invariant on their own and that's why we assigned them independent gauge parameters. Similarly, the 
	 gauge transformations associated with the $\bar{\Sm}$ supercharge are
	\begin{align}
	  \lbrack \Jm'_-,\bar{\Sm}]\l_{k} \l_{n-k} & =  \b \left( \delta_{k=0} + \delta_{n=k} \right) \l_{n}\, , & 	  
	 \\
	 [\Jm'_-,\bar{\Sm}]\l_k \bar{\Fm}_{n-k} & = \b \delta_{k=0} \bar{\Fm}_n\, ,  & [\Jm'_-,\bar{\Sm}]\bar{\Fm}_k \l_{n-k} & =  -\b  
	 \delta_{n=k} \bar{\Fm}_n\, , 
	 \\
	 [\Jm'_-,\bar{\Sm}]\l_k Q_{n-k} & = \b \delta_{k=0} Q_n\, ,  & [\Jm'_-,\bar{\Sm}]\tilde{Q}_k \l_{n-k} & = -\b 
	 \delta_{n=k}\tilde{Q}_n \, , 
	 \\
	 [\Jm'_-,\bar{\Sm}]\bar{\p}_k \l_{n-k} & =  \b  \delta_{n=k} 
	 \bar{\p}_n\, ,  & [\Jm'_-,\bar{\Sm}]\l_k \bar{\pt}_{n-k} & =  \b  \delta_{k=0} 
	 \bar{\pt}_n\, ,
	\end{align}
	and the action of $\bar{\Sm}$ consistent with them is
	\begin{align}
	\label{Sactfirst}
	  \bar{\Sm} \l_k \l_{n-k} & =  \b \l_{n+1}\, ,
	 \\
	  \Sm \l_k \bar{\Fm}_{n-k} & =  \b \frac{(n-k+2)(n-k+1)}{(n+3)(n+2)}\bar{\Fm}_{n+1}\, , 
	 \\
	  \Sm \bar{\Fm}_k \l_{n-k} & =  -\b 
	 \frac{(k+2)(k+1)}{(n+3)(n+2)}\bar{\Fm}_{n+1}\, , 
	 \\
	  \bar{\Sm} \l_k Q_{n-k} & =  \b Q_{n+1}\, , 
	  \\
	 \bar{\Sm} \tilde{Q}_k \l_{n-k} & =   -\b \tilde{Q}_{n+1}\, ,
	 \\
	  \bar{\Sm} \bar{\p}_k \l_{n-k} & =  \b \frac{k+1}{n+2}\bar{\p}_{n+1}\, , 
	  \\ 
	 \bar{\Sm} \l_k \bar{\pt}_{n-k} & =  \b \frac{n-k+1}{n+2}\bar{\pt}_{n+1}\, ,
	 \\
	 \label{Sactnlast}
	 \Sm Q_{k}^{i} \bar{\p}_{n-k\, i} & =  \b' \frac{n-k+1}{(n+2)(n+1)}\bar{\Fm}_{n}\, , 
	 \\
	 \label{Sactlast}
	 \Sm \bar{\pt}_{k}^{\tilde{\imath}} \tilde{Q}_{n-k\, \tilde{\imath}} & =  \b'' \frac{k+1}{(n+2)(n+1)}\bar{\Fm}_{n}\, .
	 \end{align}
	 As  before, the terms with $\b'$ and $\b''$ are invariant on their own.
	 Note that the action of $\bar{\Sm}$ on $Q_{k} \bar{\p}_{n-k}$ and $\bar{\pt}_{k\, } \tilde{Q}_{n-
	 k}$  is non-zero only for the flavor-contracted combinations. Indeed the action on the
	  gauge-contracted combinations would have 
	 to  give a single letter with open flavor indices which is impossible.
	 Now we impose the hermiticity condition
	\be 
	\bar{\Qm}^{\dag}  =  \bar{\Sm} \, ,
	\ee
	which implies  the following reality constraints for the undetermined coefficients:
	\begin{equation}
	\a  = \b^*\, , \quad \a'  =	\b'^*\, , \quad \a''  = \b''^*\, .
	\end{equation}
	
	\subsection{The Hamiltonian as a sum of projectors}
	
	Having determined the $O(g)$ action of ${\bar{\Qm}}(g)$ and ${\bar{\Sm}}(g)$, the one-loop Hamiltonian is easily obtained from
	\be 
	\label{algH}
	H' = 2 \{\bar{\Sm},\bar{\Qm}\} \, .
	\ee
	This result suffers  from a certain gauge ambiguity, as we may redefine
	\be \label{gaugeamb}
	H'_{\ell, \ell +1} \to H'_{\ell, \ell +1} - K_\ell + K_{\ell +1} \, ,
	\ee
	where  $H'_{\ell, \ell +1}$ is the
 Hamiltonian density at sites $(\ell, \ell +1)$ and $K_\ell$ an arbitrary local operator at site $\ell$.
	The total Hamiltonian  $H' = \sum_\ell H'_{\ell, \ell +1}$ is guaranteed to be 
	invariant under $SU(1,1)$. As in \cite{Liendo:2011xb}, we
	 fix the gauge ambiguity (\ref{gaugeamb}) by demanding the stronger condition that the Hamiltonian  density  be $SU(1,1)$ invariant as well.
	We can then write $H'_{\ell, \ell +1}$ as 
a sum of projectors onto the $SU(1,1)$ irreps of the two-site state space, with coefficients
determined by explicit evaluation on the primary of each module.
 The uplifting
procedure is straightforward: one writes the full Hamiltonian
as a sum over $SU(2,2|2)$  projectors and fixes the coefficients
by comparison with the $SU(1,1)$ subsector, as each $SU(1,1)$ primary
 is also a $SU(2,2|2)$ descendant. We simply quote the results in the various subspaces.
 	
	\subsubsection{$\Vm \times \Vm$ and $\bar{\Vm} \times \bar{\Vm}$}
	
	We find
	\be 
	H_{12}= 0 \times \Pm_{\bar{\Bm}_{2(0,0)}}+2|\a|^2\sum_{q=1}^{\infty}2h(q)\Pm_{(\frac{q+1}{2},\frac{q-2}{2})}\, ,
	\ee
	for $\Vm \times \Vm$ and
	\be 
	H_{12}= 0 \times \Pm_{\Bm_{-2(0,0)}}+2(-2|\a|^2+\frac{|\a'|^2}{2}+\frac{|\a''|^2}{2})\sum_{q=1}^{\infty}2h(q)\Pm_{(\frac{q-2}
	{2},\frac{q+1}{2})}\, ,
	\ee
	for $\bar{\Vm} \times \bar{\Vm}$. CPT invariance implies that these expressions should be identical, which imposes an extra restriction on 
	$|\a'|^2$ and $|\a''|^2$ namely,
	\be 
	|\a'|^2 + |\a''|^2=6|\a|^2\, .
	\ee
	Now, $\a'$ and $\a''$ are parameters associated with ($\Xm,\bar{\Xm}$) and ($\tilde{\Xm},\bar{\tilde{\Xm}}$) repectively. Parity 
	(which is a symmetry of the theory, see section \ref{Tables}) interchanges the two, in order to have parity invariant 
	Hamiltonian we need to set
	\bea 
	\a' & = & \sqrt{3}e^{i \theta}\a\,,
	\\
	\a''& = & -\sqrt{3}e^{i \theta}\a\, ,
	\eea
	where $\theta$ is an arbitrary phase, which can be set to zero by  a similarity transformation.
	
	\subsubsection{ $\Vm \times \Xm$, $\bar{\Xm} \times \bar{\Vm}$, $\bar{\Vm} \times \Xm$ and $\bar{\Xm} \times \Vm$}
	
	The Hamiltonian in these subspaces is\footnote{Some of the modules with low $q$ are not present in the subsector so
the corresponding	coefficients are not determined by the algebraic constraints.  However, these coefficients can be fixed by invoking CPT.}
	\begin{align}
	 \Vm \times \Xm & =  2|\a|^2\sum_{q=0}(h(q+1)+h(q)-\frac{1}{2})\Pm_{(\frac{q+1}{2},\frac{q-1}{2})}=\tilde{\Xm} \times \Vm\, ,  
	\\
	 \bar{\Xm} \times \bar{\Vm} & = 2|\a|^2 \sum_{q=0}(h(q+1)+h(q)-\frac{1}{2})\Pm_{(\frac{q-1}{2},\frac{q+1}{2})}=\bar{\Vm} \times 
	 \bar{\tilde{\Xm}}\, , 
	\\
	 \bar{\Vm} \times \Xm & =  2|\a|^2\sum_{q=0}(h(q+2)+h(q)-\frac{1}{2})\Pm_{(\frac{q}{2},\frac{q+1}{2})}\,=\tilde{\Xm} \times 
	 \bar{\Vm}\, , 
	\\
	 \bar{\Xm} \times \Vm & = 2|\a|^2 \sum_{q=0}(h(q+2)+h(q)-\frac{1}{2})\Pm_{(\frac{q+1}{2},\frac{q}{2})}= \Vm \times 
	 \bar{\tilde{\Xm}} \, .
	\end{align}
	\subsubsection{$\bar{\Xm} \times \Xm$, $\tilde{\Xm} \times \Xm$, $\tilde{\Xm} \times 
	 \bar{\tilde{\Xm}}$ and $\bar{\Xm} \times 
	 \bar{\tilde{\Xm}}$ (gauge contracted)}
	Here we find
	\begin{align}
	\label{XbX}
	 \bar{\Xm} \times \Xm & =  2|\a|^2\sum_{q=0}(h(q+1)+h(q)-1)\Pm_{(\frac{q}{2},\frac{q}{2})}=\tilde{\Xm} \times 
	 \bar{\tilde{\Xm}}\, ,  
	 \\
	 \label{XtX}	 	 
	 \tilde{\Xm} \times \Xm & = 2|\a|^2 \sum_{q=-1}(2h(q+1)-1)\Pm_{(\frac{q+1}{2},\frac{q}{2})}\, , 
	 \\	
	  \label{XbXtb} 	 
	 \bar{\Xm} \times 
	 \bar{\tilde{\Xm}} & = 2|\a|^2 \sum_{q=-1}(2h(q+1)-1)\Pm_{(\frac{q}{2},\frac{q+1}{2})}\, .
	\end{align}
	\subsubsection{\textbf{$ \Xm\times\bar{\Xm}$} (flavor contracted), \textbf{$ \bar{\tilde{\Xm}} \times \tilde{\Xm}$} (flavor 
	contracted), \textbf{$\Vm\times\bar{\Vm}$} and \textbf{$\bar{\Vm}\times\Vm$}.}
	This is the most involved subspace since we have  mixing between different copies
	of the same superconformal multiplets.
	The Hamiltonian acting on this subspace is a $4 \times 4$ 
	matrix,
	\begin{align}
	\label{4x4}
	H_{12} & = 3|\a|^2
	{\footnotesize
	\left(
	\begin{array}{c|c c c c c}
	 & \Xm\bar{\Xm} & \bar{\tilde{\Xm}}\tilde{\Xm} & \Vm\bar{\Vm} & \bar{\Vm}\Vm  
	\\
	\hline
	\bar{\Xm}\Xm  &1 &  -1& 0 &0 \\
	\tilde{\Xm}\bar{\tilde{\Xm}} & -1 & 1 & 0 &0 \\
	\bar{\Vm}\Vm & 0 & 0 & 0 &0 \\
	\Vm\bar{\Vm}  & 0 & 0 & 0 &0 \\
	\end{array} \right)\Pm_{(0,0)} } \, 
	\\
	\nn
	&  + {\footnotesize 2|\a|^2\sum_{q=1}^{\infty} 
	\left(
	\begin{array}{c|c c c c c}
	 & \Xm\bar{\Xm} & \bar{\tilde{\Xm}}\tilde{\Xm} & \Vm\bar{\Vm} & \bar{\Vm}\Vm  
	\\
	\hline
	\bar{\Xm}\Xm   & 0 & 0 & \sqrt{3}e^{-i\theta} & -\sqrt{3}\frac{e^{-i\theta}}{q+1} \\
	\tilde{\Xm}\bar{\tilde{\Xm}} & 0 & 0 & -\sqrt{3}\frac{e^{-i\theta}}{q+1} & \sqrt{3}e^{-i\theta}\\
	\bar{\Vm}\Vm & \sqrt{3}\frac{e^{i\theta}}{q(q+2)} &-\sqrt{3}\frac{e^{i\theta}}{q(q+1)(q+2)} 
	& h(q+2)+ h(q-1) & \frac{2}{q(q+1)(q+2)} \nn \\
	\Vm\bar{\Vm} &-\sqrt{3}\frac{e^{i\theta}}{q(q+1)(q+2)} &\sqrt{3}\frac{e^{i\theta}}{q(q+2)} 
	&  \frac{2}{q(q+1)(q+2)} & h(q+2)+ h(q-1)
	\end{array} \right)\Pm_{(\frac{q}{2},\frac{q}{2} )}} \, .
	\end{align}

	\subsection{Scalar Sector}
	
Let us compare our results with the scalar 
	sector computation of \cite{Poland:2011kg}. Apart from providing
	a check of our procedure, this comparison
	allows us to fix the overall normalization of the Hamiltonian in terms of the gauge theory 't Hooft coupling.
	The action of the Hamiltonian for the gauge contracted $Q$ and $\tilde{Q}$ pairs can be   
	obtained from (\ref{XbX}) at $q=0$ and (\ref{XtX},\ref{XbXtb}) at $q=-1$,
	\be 
	H_{12}=2|\a|^2
	\left(
	\begin{array}{c| c c c c}
	& \bar Q Q & \bar Q \bar{\tilde{Q}} & \tilde Q Q & \tilde Q \bar{\tilde{Q}}
	\\
	\hline	
	Q \bar Q & 0 &  & &  \\
	Q \tilde Q & & -1 & & \\
	\bar{\tilde Q} \bar{Q} &  &  & -1 &  \\
	\bar{\tilde Q} \tilde Q&  &  &  & 0 
	\end{array} \right)\, ,
	\ee
	in perfect agreement with equation (3.6) of \cite{Poland:2011kg} provided we identify $2|\a|^2=\l$.
	For the flavor contracted pairs we obtain from the first matrix in (\ref{4x4}),
	\be 
	H_{12}= 3|\a|^2
	\left(
	\begin{array}{c| c c}
	& Q \bar Q & \bar{\tilde{Q}} \tilde{Q}
	\\
	\hline
	\bar Q Q & 1 & -1  \\
	\tilde{Q} \bar{\tilde{Q}} & -1 & 1 \\
	\end{array} \right)\, ,
	\ee
	again in agreement with \cite{Poland:2011kg}. It is interesting   that the value 
	of the transverse magnetic field for the Ising spin chain
	in the scalar sector, namely $h_{\textrm{Ising}} = N_f/N_c = 3$ \cite{Poland:2011kg},
	 turns out to be determined by superconformal symmetry alone.

	\subsection{Open Chain}
	
	The extension of the previous results to an open chain with adjoint or bi-fundamental flavor indices is straightforward. 
	To obtain the full Hamiltonian we start considering the open chain states of the $SU(1,1)$ subsector,
	\be 
	\label{openstates}
	\bar \psi_{a i} \ldots Q^{a j}\, , \quad \bar{\psi}_{a i} \ldots \bar{\tilde{\psi}}^{a \tilde \jmath}\, ,
	\quad \tilde Q_{a \tilde \imath} \ldots  Q^{a j}\, , 
	\quad \tilde Q_{a \tilde \imath} \ldots \bar{\tilde{\psi}}^{a \tilde \jmath}\, ,
	\ee	
	where $_i^{\ph{ab}j}$, $_\itil^{\ph{ab}\jtil}$  are projected to its adjoint irreducible component  	
	(the singlet is subtracted). One easily check that the gauge transformations (\ref{Qactfirst}--\ref{Qactlast}) and  (\ref{Sactfirst}--\ref{Sactlast})  leave
	invariant  open chains of this form. The uplift to the full Hamiltonian works just as for the closed chain.
	The upshot is that the Hamiltonian density derived for the closed chain applies with no modification
	to the open chains.
	
\subsection{The Harmonic Action}
\label{harmonicS}

In this section we present an explicit oscillator form of the Hamiltonian analogous to Beisert's harmonic action \cite{Beisert:2003jj} for ${\cal N}=4$ SYM.

\subsubsection{\textbf{$\Vm \times \Vm$} and \textbf{$\bar{\Vm} \times \bar{\Vm}$}.}

For states in these two subspaces the action of the Hamiltonian is identical with that of $\Nm=4$ SYM and of $\Nm=2$ SCQCD. General states in $\Vm \times \Vm$ and $\bar{\Vm} \times \bar{\Vm}$ can be written as 
\bea 
|s_1,...,s_n;A\rangle_{\text{\tiny{$\Vm\times\Vm$}}} & = & A^\dag_{s_1 A_1}... A^\dag_{s_n A_n}|0\rangle \otimes |0\rangle\, ,
\\
|s_1,...,s_n;A\rangle_{\text{\tiny{$\bar{\Vm}\times\bar{\Vm}$}}} & = & A^\dag_{s_1 A_1}... A^\dag_{s_n A_n}|\db_3\rangle \otimes |\db_3\rangle\, ,
\eea
where $A^\dag_A=(\ab^\dag_{\a},\bb^\dag_{\ad},\cb^\dag)$ and $s_i=1,2$ indicates in which site the oscillator sits. The action of the Hamiltonian on this state does not change the number of oscillators but merely shifts them from site 1 to site 2 (or vice versa) in all possible combinations. This can be written as
\bea
H_{12}|s_1,...,s_n;A\rangle_{\text{\tiny{$\Vm\times\Vm$}}} & = & \sum_{s'_1,...,s'_n}c_{n,n_{12},n_{21}}\delta_{C_1,0} \delta_{C_2,0}|s'_1,...,s'_n;A\rangle_{\text{\tiny{$\Vm\times\Vm$}}} \, ,
\\
H_{12}|s_1,...,s_n;A\rangle_{\text{\tiny{$\bar{\Vm}\times\bar{\Vm}$}}} & = & \sum_{s'_1,...,s'_n}c_{n,n_{12},n_{21}}\delta_{C_1,0} \delta_{C_2,0}|s'_1,...,s'_n;A\rangle_{\text{\tiny{$\bar{\Vm}\times\bar{\Vm}$}}} \, ,
\eea
where the Kronecker deltas project onto states with zero central charge and $n_{ij}$ counts the number of oscillators moving from site $i$ to site $j$. The explicit formula for the function $c_{n,n_{12},n_{21}}$ is
\be 
c_{n,n_{12},n_{21}}=(-1)^{1+n_{12}n_{21}}\frac{\Gamma(\frac{1}{2}(n_{12}+n_{21}))\Gamma(1+\frac{1}{2}(n-n_{12}-n_{21}))}{\Gamma(1+\frac{1}{2}n)}\, ,
\ee
with $c_{n,0,0}=h(\frac{n}{2})$. In \cite{Beisert:2003jj} it was proven that this function is a superconformal invariant and that it has the appropriate eigenvalues when acting on the $\hat{\Cm}_{(\frac{q+1}{2},\frac{q-2}{2})}$ and $\hat{\Cm}_{(\frac{q-2}{2},\frac{q+1}{2})}$ modules, namely
\bea
H_{12}\hat{\Cm}_{(\frac{q+1}{2},\frac{q-2}{2})} & = & 2h(q)\hat{\Cm}_{(\frac{q+1}{2},\frac{q-2}{2})}\,,
\\
H_{12}\hat{\Cm}_{(\frac{q-2}{2},\frac{q+1}{2})} & = & 2h(q)\hat{\Cm}_{(\frac{q-2}{2},\frac{q+1}{2})}\,.
\eea

\subsubsection{\textbf{$\Vm \times \Xm$}\,, \textbf{$\tilde{\Xm} \times \Vm$}\,, \textbf{$\bar{\Xm} \times \bar{\Vm}$} and \text{$\bar{\Vm} \times \bar{\tilde{\Xm}}$}.}

General states in these four subspaces can be written as
\bea
|s_1,...,s_n;A\rangle_{\text{\tiny{$\Vm\times\Xm$}}} & = & A^\dag_{s_1 A_1}... A^\dag_{s_n A_n}|0\rangle \otimes |\db_1\rangle\,,
\\
|s_1,...,s_n;A\rangle_{\text{\tiny{$\tilde{\Xm}\times\Vm$}}} & = & A^\dag_{s_1 A_1}... A^\dag_{s_n A_n}|\tilde{\db}_1\rangle \otimes |0\rangle\,,
\\
|s_1,...,s_n;A\rangle_{\text{\tiny{$\bar{\Xm}\times\bar{\Vm}$}}} & = & A^\dag_{s_1 A_1}... A^\dag_{s_n A_n}|\db_1\db_2\rangle \otimes |\db_3\rangle\,,
\\
|s_1,...,s_n;A\rangle_{\text{\tiny{$\bar{\Vm}\times\bar{\tilde{\Xm}}$}}} & = & A^\dag_{s_1 A_1}... A^\dag_{s_n A_n}|\db_3\rangle \otimes |\tilde{\db}_1\tilde{\db}_2\rangle\,,
\eea
where $|\db_i\rangle = \db^\dag_i |0\rangle$ \footnote{The tilde in some of the $\db$ oscillators is just a reminder that we are looking at the $\tilde{\Xm}$ multiplet or its conjugate.}.
The action of $H_{12}$ for all these four subspaces is given by \footnote{To simplify the notation we will omit the Kronecker deltas  $\delta_{C_1,0} \delta_{C_2,0}$.}
\be
H_{12}|s_1,...,s_n;A\rangle = \sum_{s'_1,...,s'_n}c_{n+1,n_{12},n_{21}}|s_1,...,s_n;A\rangle-\frac{1}{2}|s_1,...,s_n;A\rangle\,.
\ee
Invariance under the superconformal group is guaranteed because we are using the $c_{n,n_{12},n_{21}}$ constants. We need only to check that this expression has the correct eigenvalues when acting on the corresponding $\Nm=1$ primaries (see Appendix \ref{PrimariesA}), which can be easily done using an algebra software like \texttt{Mathematica}.

\subsubsection{\textbf{$\bar{\Vm} \times \Xm$}\,, \textbf{$\tilde{\Xm} \times \bar{\Vm}$}\,, \textbf{$\bar{\Xm} \times \Vm$} and \text{$\Vm \times \bar{\tilde{\Xm}}$}. }

In this case we have
\bea
|s_1,...,s_n;A\rangle_{\text{\tiny{$\bar{\Vm}\times\Xm$}}} & = & A^\dag_{s_1 A_1}... A^\dag_{s_n A_n}|\db_3\rangle \otimes |\db_1\rangle\,,
\\
|s_1,...,s_n;A\rangle_{\text{\tiny{$\tilde{\Xm}\times\bar{\Vm}$}}} & = & A^\dag_{s_1 A_1}... A^\dag_{s_n A_n}|\tilde{\db}_1\rangle \otimes |\db_3\rangle\,,
\\
|s_1,...,s_n;A\rangle_{\text{\tiny{$\bar{\Xm}\times\Vm$}}} & = & A^\dag_{s_1 A_1}... A^\dag_{s_n A_n}|\db_1\db_2\rangle \otimes |0\rangle\,,
\\
|s_1,...,s_n;A\rangle_{\text{\tiny{$\Vm\times\bar{\tilde{\Xm}}$}}} & = & A^\dag_{s_1 A_1}... A^\dag_{s_n A_n}|0\rangle \otimes |\tilde{\db}_1\tilde{\db}_2\rangle\,,
\eea
and the action of $H_{12}$ reads
\be
H_{12}|s_1,...,s_n;A\rangle = \sum_{s'_1,...,s'_n}c_{n+2,n_{12},n_{21}}|s_1,...,s_n;A\rangle-\frac{1}{2}|s_1,...,s_n;A\rangle\,.
\ee

\subsubsection{\textbf{$\bar{\Xm} \times \Xm$} and \textbf{$\tilde{\Xm} \times \bar{\tilde{\Xm}}$} (gauge contracted).}

The states are given by
\bea
|s_1,...,s_n;A\rangle_{\text{\tiny{$\bar{\Xm}\times\Xm$}}} & = & A^\dag_{s_1 A_1}... A^\dag_{s_n A_n}|\db_1 \db_2\rangle \otimes |\db_1\rangle\,,
\\
|s_1,...,s_n;A\rangle_{\text{\tiny{$\tilde{\Xm}\times\bar{\tilde{\Xm}}$}}} & = & A^\dag_{s_1 A_1}... A^\dag_{s_n A_n}|\tilde{\db}_1\rangle \otimes |\tilde{\db}_1\tilde{\db}_2\rangle\,,
\eea
and the action of $H_{12}$ is
\be
H_{12}|s_1,...,s_n;A\rangle = \sum_{s'_1,...,s'_n}c_{n+1,n_{12},n_{21}}|s_1,...,s_n;A\rangle-|s_1,...,s_n;A\rangle\,.
\ee
\subsubsection{\textbf{$\tilde{\Xm} \times \Xm$} and \text{$\bar{\Xm} \times \bar{\tilde{\Xm}}$}.}

The states are given by
\bea
|s_1,...,s_n;A\rangle_{\text{\tiny{$\tilde{\Xm} \times \Xm$}}} & = & A^\dag_{s_1 A_1}... A^\dag_{s_n A_n}|\tilde{\db}_1\rangle \otimes |\db_1\rangle\,,
\\
|s_1,...,s_n;A\rangle_{\text{\tiny{$\bar{\Xm} \times \bar{\tilde{\Xm}}$}}} & = & A^\dag_{s_1 A_1}... A^\dag_{s_n A_n}|\db_1\db_2\rangle \otimes |\tilde{\db}_1\tilde{\db}_2\rangle\,,
\eea
and the action of $H_{12}$ is
\be
H_{12}|s_1,...,s_n;A\rangle = \sum_{s'_1,...,s'_n}\left(c_{n+2,n_{12},n_{21}}-c_{n+2,n_{12}+1,n_{21}+1}\right)|s_1,...,s_n;A\rangle-|s_1,...,s_n;A\rangle\,.
\ee

\subsubsection{\textbf{$ \Xm\times\bar{\Xm}$} (flavor contracted), \textbf{$ \bar{\tilde{\Xm}} \times \tilde{\Xm}$} (flavor contracted), \textbf{$\Vm\times\bar{\Vm}$} and \textbf{$\bar{\Vm}\times\Vm$}.}

The states are
\bea
|s_1,...,s_n;A\rangle_{\text{\tiny{$\Xm \times \bar{\Xm}$}}} & = & A^\dag_{s_1 A_1}... A^\dag_{s_n A_n}|\db_1\rangle \otimes |\db_1\db_2\rangle\,,
\\
|s_1,...,s_n;A\rangle_{\text{\tiny{$\bar{\tilde{\Xm}} \times \tilde{\Xm}$}}} & = & A^\dag_{s_1 A_1}... A^\dag_{s_n A_n}|\tilde{\db}_1\tilde{\db}_2\rangle \otimes |\tilde{\db}_1\rangle\,,
\\
|s_1,...,s_n;A\rangle_{\text{\tiny{$\Vm \times \bar{\Vm}$}}} & = & A^\dag_{s_1 A_1}... A^\dag_{s_n A_n}|0\rangle \otimes |\db_3\rangle\,,
\\
|s_1,...,s_n;A\rangle_{\text{\tiny{$\bar{\Vm} \times \Vm$}}} & = & A^\dag_{s_1 A_1}... A^\dag_{s_n A_n}|\db_3\rangle \otimes |0\rangle\,.
\eea
The action of $H_{12}$ is given by
\bea
\nn
H_{12}|s_1,...,s_n;A\rangle_{\text{\tiny{$\Xm\times\bar{\Xm}$}}} & = &  \sum_{s'_1,...,s'_n} c_{n+3,n_{12}+1,n_{21}+1}
|s_1,...,s_n;A\rangle_{\text{\tiny{$\Xm\times\bar{\Xm}$}}}
\\
\nn
& & + 3 \sum_{s'_1,...,s'_n}c_{n+3,n_{12}+1,n_{21}+2}|s_1,...,s_n;A\rangle_{\text{\tiny{$\bar{\tilde{\Xm}}\times \tilde{\Xm}$}}}
\\
\nn
& & + \sqrt{3} e^{-i\theta}\sum_{s'_1,...,s'_n}c_{n+3,n_{12},n_{21}+1}|s_1,...,s_n;A\rangle_{\text{\tiny{$\Vm\times \bar{\Vm}$}}}
\\
& & + \sqrt{3} e^{-i\theta}\sum_{s'_1,...,s'_n}c_{n+3,n_{12}+1,n_{21}+1}|s_1,...,s_n;A\rangle_{\text{\tiny{$\bar{\Vm}\times \Vm$}}}\, ,
\eea
\bea
\nn
H_{12}|s_1,...,s_n;A\rangle_{\text{\tiny{$\bar{\tilde{\Xm}}\times \tilde{\Xm}$}}} & = &  \sum_{s'_1,...,s'_n} c_{n+3,n_{12}+1,n_{21}+1}
|s_1,...,s_n;A\rangle_{\text{\tiny{$\bar{\tilde{\Xm}}\times \tilde{\Xm}$}}}
\\
\nn
& & - 3\sum_{s'_1,...,s'_n}c_{n+3,n_{12}+1,n_{21}+2}|s_1,...,s_n;A\rangle_{\text{\tiny{$\Xm\times\bar{\Xm}$}}}
\\
\nn
& & + \sqrt{3} e^{-i\theta}\sum_{s'_1,...,s'_n}c_{n+3,n_{12}+1,n_{21}+1}|s_1,...,s_n;A\rangle_{\text{\tiny{$\Vm\times \bar{\Vm}$}}}
\\
& & + \sqrt{3} e^{-i\theta}\sum_{s'_1,...,s'_n}c_{n+3,n_{12}+1,n_{21}}|s_1,...,s_n;A\rangle_{\text{\tiny{$\bar{\Vm}\times \Vm$}}}\, ,
\eea
\bea
\nn
H_{12}|s_1,...,s_n;A\rangle_{\text{\tiny{$\Vm\times \bar{\Vm}$}}} & = &  \sum_{s'_1,...,s'_n} c_{n+3,n_{12},n_{21}}
|s_1,...,s_n;A\rangle_{\text{\tiny{$\Vm\times \bar{\Vm}$}}}
\\
\nn
& & + \sum_{s'_1,...,s'_n}c_{n+3,n_{12}+2,n_{21}+1}|s_1,...,s_n;A\rangle_{\text{\tiny{$\bar{\Vm}\times \Vm$}}}
\\
\nn
& & + \sqrt{3} e^{i\theta}\sum_{s'_1,...,s'_n}c_{n+3,n_{12}+1,n_{21}}|s_1,...,s_n;A\rangle_{\text{\tiny{$\Xm\times\bar{\Xm}$}}}
\\
& & - \sqrt{3} e^{i\theta}\sum_{s'_1,...,s'_n}c_{n+3,n_{12},n_{21}+2}|s_1,...,s_n;A\rangle_{\text{\tiny{$\bar{\tilde{\Xm}}\times \tilde{\Xm}$}}}\, ,
\eea
and
\bea
\nn
H_{12}|s_1,...,s_n;A\rangle_{\text{\tiny{$\bar{\Vm}\times \Vm$}}} & = &  \sum_{s'_1,...,s'_n} c_{n+3,n_{12},n_{21}}
|s_1,...,s_n;A\rangle_{\text{\tiny{$\bar{\Vm}\times \Vm$}}}
\\
\nn
& & - \sum_{s'_1,...,s'_n}c_{n+3,n_{12}+2,n_{21}+1}|s_1,...,s_n;A\rangle_{\text{\tiny{$\Vm\times \bar{\Vm}$}}}
\\
\nn
& & - \sqrt{3} e^{i\theta}\sum_{s'_1,...,s'_n}c_{n+3,n_{12}+2,n_{21}}|s_1,...,s_n;A\rangle_{\text{\tiny{$\Xm\times\bar{\Xm}$}}}
\\
& & + \sqrt{3} e^{i\theta}\sum_{s'_1,...,s'_n}c_{n+3,n_{12},n_{21}+1}|s_1,...,s_n;A\rangle_{\text{\tiny{$\bar{\tilde{\Xm}}\times \tilde{\Xm}$}}}\,.
\eea

\section{Spectral analysis}
\label{Tables}

Spectral studies in planar $\Nm=4$ SYM have shown the systematic presence of degenerate pairs of states
of opposite ``parity'', where parity is the $\mathbb{Z}_2$ symmetry associated with complex conjugation  of the $SU(N)$ gauge group \cite{Beisert:2003tq,Beisert:2003ys,Zwiebel:2005er,Beisert:2004ry}.
These degeneracies persist at higher loops, but are lifted by non-planar corrections. This phenomenon is naturally explained by 
the integrable structures of planar ${\cal N}=4$ SYM: the theory admits
higher conserved charges  that are parity-odd and map the degenerate eigenstates. In some models it is even
possible to prove that the existence of parity pairs is a sufficient condition for integrability \cite{Beisert:2004ry}.

The upshot is that in ${\cal N}=4$ SYM the existence of parity pairs is one of the many pieces of evidence for the complete integrability
of the theory. With this precedent in mind, 
we  can look forward to a similar analysis in ${\cal N}=2$ SCQCD and in ${\cal N}=1$ SQCD.
In this section we determine the low-lying  spectrum of the one-loop dilation operator
of both theories, in the closed non-compact subsectors that were used to uplift the full Hamiltonian.

\subsection{$\Nm=2$ SCQCD}

We start our analysis with $\Nm=2$ SCQCD and with the more general  
quiver theory that interpolates between the $\mathbb{Z}_2$ orbifold of ${\cal N}=4$ SYM and
$\Nm=2$ SCQCD. 
For background material  we refer the reader to \cite{Gadde:2010zi, Gadde:2010ku, Liendo:2011xb} where this model was studied extensively.

\subsubsection{Parity}

The first thing we need to do is define a meaningful parity operation. We will take $\Nm=4$ SYM as our starting point where parity amounts to conjugation of the $SU(N)$ gauge group. Under parity, the Lie algebra generators transform as
\be 
T^a_{\ph{a}b} \rightarrow -(T^{a}_{\ph{a}b})^* = -T^b_{\ph{b}a}\, ,
\ee
where we have used hermiticity to trade conjugation by transposition. 

Now, as reviewed in \cite{Gadde:2009dj,Gadde:2010zi} $\Nm=2$ SCQCD can be thought of as a limit of a two-parameter $(g,\check{g})$ quiver theory with gauge
group  $SU(N_c) \times SU(N_{\check c})$ (with $N_{\check c} \equiv N_c$): one has 
 $\Nm=2$ SCQCD at $\check{g}=0$ and the $\mathbb{Z}_2$ orbifold of $\Nm=4$ SYM at $g=\check{g}$. Starting from $\Nm=4$ SYM with gauge group $SU(2N_c)$
 the $\mathbb{Z}_2$ orbifold theory is obtained by
the projection
 \begin{eqnarray}  \label{survive}
A_{\a \dot \a} & = &   \left( \begin{array}{cc}
A_{\a \dot \a\, b}^{a} & 0 \\
0 &  \topp A_{\a \dot \a\, \topp b}^{\topp a} \end{array} \right)\, , \quad
Z  =   \left( \begin{array}{cc}
 \f_{\mbox{ } \mbox{ }b}^{a} & 0 \\
0 &  \topp{\f}_{\mbox{ } \mbox{ } \topp b}^{\topp a} \end{array} \right)\, , 
\\
 \lambda_{\II}& =&  \left( \begin{array}{cc}
 \lambda_{\II b}^{a} & 0 \\
0 &  \topp{\la}_{\II \topp b}^{\topp a} \end{array} \right)\, ,\quad
 \lambda_{\hat{\II}}  =   \left( \begin{array}{cc}
0 &  \epsilon_{\hat{\II} \hat{\JJ}}\psi^{a\hat{\JJ}}_{\ph{a} \topp a} \\
 \epsilon_{\hat{\II} \hat{\JJ}}\tilde{\psi}^{\topp b\hat{\JJ}}_{\ph{b} b} & 0 \end{array} \right)\, ,
 \label{fermident}
\\
 \mathcal{X}_{\II \IIh} & = &  \left( \begin{array}{cc}
0 & Q_{\II \IIh \topp a}^{\mbox{ }a} \\
- \epsilon_{\II \JJ} \epsilon_{\hat{\II} \hat{\JJ}} \bar{Q}_{\mbox{ } \mbox{ }b}^{\topp b \hat{\JJ} \JJ} & 0 \end{array} \right) 
  \, ,
 \end{eqnarray}
where  $\Im, \hat \Im=1,2$. 
The parity operation  described above implies the following transformations. For the fields in the vector multiplets,
\begin{equation}
\begin{split}
A^a_{\a \dot \a\, b}  \lra -A^b_{\a \dot \a\, a}  \quad  \l^a_{\Im b}  \lra -\l^b_{\Im a} \quad  \f^a_{\ph{a} b}  \lra -\f^b_{\ph{b}a}\, ,\\
\check A^{\check  a}_{\a \dot \a\, \check b}  \lra -\check A^{\check b}_{\a \dot \a\,\check  a}  \quad \check \l^{\check a}_{\Im \check  b}  \lra -\check \l^{\check b}_{\Im \check a} \quad \check \f^{\check a}_{\ph{a}\check b}  \lra -\check \f^{\check b}_{\ph{b} \check a}\, ,
\end{split}
\end{equation}
and analogous expressions for the conjugate fields.
For the fields in the hypermultiplets,
\begin{equation}
\begin{split}
\psi^a_{\hat \Im \check b}  \lra -\tilde \psi^{\check b}_{\hat \Im a}  \quad Q^a_{\Im \hat \Im \check b}  \lra \bar Q^{\check b}_{\Im \hat \Im a}\, , 
\end{split}
\end{equation}
and analogous expressions for the conjugate fields.
These transformations remain a symmetry also away from the orbifold point (that is, for arbitrary $(g, \check g))$,  
as can be easily checked by inspection of the Lagrangian (see {\it e.g.} \cite{Gadde:2010zi} for the explicit  expression of the Lagrangian). 
This implies that the parity operation commutes with the dilation operator to all loops. 
Its action on single-trace states is then given by
\be 
P | A_1\, .\, .\, .\, A_L \rangle = (-1)^{L+k(k+1)/2}| A_L\, .\, .\, .\, A_1 \rangle\, ,
\ee
where $k$ is the number of fermions and we replace $\psi  \lra \tilde \psi$, $\bar \psi  \lra \bar{\tilde{\psi}}$, $Q  \lra -\bar Q$.

\subsubsection{Diagonalization}

We consider the $SU(1,1) \times SU(1|1)  \times SU(1|1) \times U(1)$ subsector defined in \cite{Liendo:2011xb}.
We focus on $SU(1,1)$ primaries (since descendants have the same anomalous dimensions)
and take with no loss of generalities  $r\ge0$. Table \ref{N4} corresponds to states with maximal $r$-charge. This subsector is made exclusively out of $\{\Dm^k\bar{\l}\}$ and is therefore identical to the $SU(1,1)$ subsector used in \cite{Beisert:2004ry} to obtain the complete dilation operator of $\Nm=4$ SYM. Being a subsector of $\Nm=4$, it is integrable and, indeed, our results indicate that degenerate states with opposite parity show up consistently at each stage of the diagonalization. The notation $P_n(x)$ denotes the roots of a polynomial of order $n$, we will not write the polynomial explicitly because we are really interested in the amount of parity pairs and not in the actual values of the energies. We will denote by $P_n(x)$ all the polynomials of order $n$ we encounter, even if they are  different from each other. 
 \begin{table}[h]
 \begin{centering}
 {\footnotesize
 \begin{tabular}{|c|c|c|c|c|}
 \hline 
  $L$  & $r$  & $\Delta_0$  & $\delta \Delta^P\, [2g_{YM}^2N/\pi^2]$ \tabularnewline
 \hline
  \hline 
3   & $\frac{3}{2}$    &  7.5  & $\frac{5}{4}^\pm$ \tabularnewline
    &                  &  9.5  & $\frac{133}{96}^\pm$ \tabularnewline
    &                  &  10.5  & $\frac{761}{480}^\pm$, $\frac{761}{560}^-$ \tabularnewline
    &                  &  11.5  & $\frac{179}{120}^\pm$ \tabularnewline
 \hline
 \hline 
4   &   $2$            &  8  & $\frac{5}{4}^\pm$ \tabularnewline
    &                  &  9  & $\frac{1}{48}(73\pm\sqrt{37})^-$ \tabularnewline
    &                  &  10 & $\frac{19}{12}^\pm$,$\frac{133}{96}^\pm$ \tabularnewline
    &                  &  11 & $P_3(x)^-$, $\frac{761}{480}^\pm$ \tabularnewline
 \hline
 \hline 
 5  &  $\frac{5}{2}$   &  9.5  & $\frac{1}{48}(73\pm\sqrt{37})^+$ \tabularnewline
    &                  &  10.5 & $\frac{7}{4}^\pm$,$\frac{19}{12}^\pm$  \tabularnewline
    &                  &  11.5 & $P_3(x)^-$,  $\frac{1}{24}(43\pm\sqrt{5})^\pm$  \tabularnewline
 \hline
  \hline
   \multicolumn{4}{|c|}{Paired eigenvalues: $\sim$ 69 \%} \tabularnewline
 \hline
 \end{tabular}
 }
 \par \end{centering}
 \caption{$SU(1,1)$ primaries with maximal $r$-charge $(r = \frac{L}{2})$
  in the $SU(1,1) \times SU(1|1) \times SU(1|1) \times U(1)$ sector of ${\cal N}=2$ SCQCD.
 We have omitted the one-dimensional subspaces where there is no room for a parity pair.}
\label{N4}
 \end{table}

Being identical to the analogous $\Nm=4$ SYM sector, we cannot use the results of Table \ref{N4} as a test for integrability. The true dynamics of $\Nm=2$ SCQCD is encoded in subspaces where the $r$-charge is not maximal. For this we need states with $Q$ and $\bar{Q}$. Our results 
are presented in Table \ref{N2}. As opposed to the results of Table \ref{N4} the presence of parity pairs here is less systematic.

 \begin{table}[h]
 \begin{centering}
 {\footnotesize
 \begin{tabular}{|c|c|c|c|c|}
 \hline 
  $L$  & $r$  & $\Delta_0$  & $\delta \Delta^P\, [2g_{YM}^2N/\pi^2]$ \tabularnewline
 \hline
  \hline 
 3  &  $\frac{1}{2}$   &  4.5  & $\frac{3}{4}^-$, $\frac{3}{4}^-$, $\frac{3}{8}^+$ \tabularnewline
    &                  &  5.5  & $\frac{15}{16}^-$, $\frac{1}{24}(16\pm\sqrt{31})^-$, $\frac{1}{32}(21\pm\sqrt{57})^+$ \tabularnewline
    &                  &  6.5  & $\frac{25}{24}^-$, $\frac{25}{24}^-$, $\frac{25}{48}^+$ \tabularnewline
    &                  &       & $\frac{1}{96}(81\pm\sqrt{561})^-$, $\frac{1}{96}(83\pm\sqrt{409})^+$ \tabularnewline
 \hline
 \hline 
 4  &  $0$             &  5  & $\frac{3}{4}^\pm$, $\frac{3}{8}^-$ \tabularnewline
    &                  &  6  & $1^\pm$, $\frac{15}{16}^+$, $\frac{15}{16}^+$ \tabularnewline
    &                  &     & $\frac{1}{32}(21\pm\sqrt{57})^-$, $\frac{1}{32}(21\pm\sqrt{57})^-$ \tabularnewline
    &                  &     & $\frac{1}{24}(16\pm\sqrt{31})^+$, $\frac{5}{8}$, $0^+$ \tabularnewline
    &  $1$             &  6  & $1^-$, $\frac{15}{16}^+$, $\frac{1}{32}(21\pm\sqrt{57})^-$ \tabularnewline
    &                  &  7  & $\frac{5}{4}^\pm$, $\frac{9}{8}^-$, $\frac{25}{24}^+$ \tabularnewline
    &                  &     & $\frac{1}{16}(16\pm\sqrt{6})^+$, $\frac{1}{96}(81\pm\sqrt{561})^+$ \tabularnewline
    &                  &     & $\frac{1}{96}(83\pm\sqrt{409})^-$\tabularnewline
 \hline
 \hline 
 5  &  $\frac{1}{2}$   &  6.5  & $1^\pm$, $1^+$, $\frac{15}{16}^-$   \tabularnewline
    &                  &       & $\frac{1}{32}(21\pm\sqrt{57})^+$   \tabularnewline
 \hline
 \hline
   \multicolumn{4}{|c|}{Paired eigenvalues: 16 \%} \tabularnewline
 \hline
 \end{tabular}
 }
 \par \end{centering}
 \caption{$SU(1,1)$ primaries with $0 \leq r < \frac{L}{2}$
  in the $SU(1,1) \times SU(1|1) \times SU(1|1) \times U(1)$ sector of ${\cal N}=2$ SCQCD.
}
\label{N2}
 \end{table}

  \begin{table}[h]
 \begin{centering}
  {\footnotesize
 \begin{tabular}{|c|c|c|c|c|}
 \hline 
  $L$  & $r$  & $\Delta_0$  & $\delta \Delta^P\, [2g_{YM}^2N/\pi^2]$ \tabularnewline
 \hline
  \hline  
 3  &  $\frac{1}{2}$   &  4.5  & $\frac{1}{2}^+$, $\frac{3}{4}^-$, $\frac{3}{4}^-$, $\frac{3}{4}^-$ \tabularnewline
    &                  &       & $\frac{3}{4}^\pm$ \tabularnewline
    &                  &  5.5  & $\frac{3}{4}^-$, $\frac{7}{8}^-$, $\frac{25}{24}^-$, $\frac{1}{2}^+$\tabularnewline
    &                  &       & $\frac{15}{16}^\pm$, $\frac{1}{32}(27\pm\sqrt{57})^\pm$ \tabularnewline
    &                  &  6.5  & $\frac{3}{4}^+$ , $\frac{25}{24}^-$, $\frac{25}{24}^-$, $\frac{25}{24}^-$ \tabularnewline
    &                  &       & $\frac{5}{4}^\pm$, $\frac{15}{16}^\pm$, $\frac{25}{24}^\pm$ \tabularnewline 
    &                  &       & $\frac{1}{96}(93\pm\sqrt{249})^\pm$ \tabularnewline
 \hline
 \hline 
  4 &  $0$             &  5  & $\frac{1}{2}^-$, $\frac{1}{2}^-$, $\frac{1}{8}(5\pm\sqrt{13})^-$ \tabularnewline
    &                  &     & $\frac{3}{4}^\pm$, $\frac{3}{4}^\pm$ \tabularnewline
    &                  &  6  & $\frac{7}{8}^+$, $\frac{25}{24}^+$, $\frac{1}{2}^-$, $\frac{1}{2}^-$ \tabularnewline
    &                  &     & $\frac{1}{8}(5\pm\sqrt{5})^+$ \tabularnewline
    &                  &     & $\frac{3}{4}^\pm$, $\frac{5}{4}^\pm$, $\frac{5}{8}^\pm$, $\frac{7}{8}^\pm$ \tabularnewline
    &                  &     & $\frac{15}{16}^\pm$, $\frac{15}{16}^\pm$, $\frac{1}{4}(3\pm\sqrt{2})^\pm$\tabularnewline
    &                  &     & $\frac{1}{32}(27\pm\sqrt{57})^\pm$ , $\frac{1}{32}(27\pm\sqrt{57})^\pm$ \tabularnewline
 \hline
 \hline
   \multicolumn{4}{|c|}{Paired eigenvalues: $\sim$ 68 \%} \tabularnewline
 \hline
 \end{tabular}
 }
 \par \end{centering}
 \caption{
 $SU(1,1)$ primaries with $0 \leq r < \frac{L}{2}$
  in the $SU(1,1) \times SU(1|1) \times SU(1|1) \times U(1)$ sector of the orbifold theory $(\check g = g)$.
We have restricted the diagonalization to $SU(2)_L$ singlets.}
\label{Z2}
 \end{table} 
 \begin{table}[h]
 \begin{centering}
 {\footnotesize
 \begin{tabular}{|c|c|c|c|c|c|c|c|}
 \hline 
$L$ & $r$ & $\Delta_0$ & $\kappa=1$   &  $\kappa=0.7$   & $\kappa=0.3$  & $\kappa=0$ & SCQCD \tabularnewline
 \hline
 \hline
 3 & $\frac{1}{2}$ & 5.5 & $\frac{1}{32}(27+\sqrt{57})$ & $0.97$ &  $0.94$  & $\frac{15}{16}$              & Yes \tabularnewline
   &               &     & $\frac{1}{32}(27+\sqrt{57})$ & $0.95$ &  $0.90$  & $\frac{1}{32}(27+\sqrt{57})$ & Yes \tabularnewline
 \hline
 3 & $\frac{1}{2}$ & 5.5 & $\frac{1}{32}(27-\sqrt{57})$ & $0.39$ &  $0.08$  &  $0$          & No\tabularnewline
   &               &     & $\frac{1}{32}(27-\sqrt{57})$ & $0.47$ &  $0.30$  & $\frac{1}{4}$ & No \tabularnewline
 \hline
 3 & $\frac{1}{2}$ & 6.5 & $\frac{5}{4}$ & $1.12$ &  $1.09$  & $\frac{1}{96}(81+\sqrt{561})$ & Yes \tabularnewline
   &               &     & $\frac{5}{4}$ & $1.11$ &  $1.08$  & $\frac{1}{96}(83+\sqrt{409})$ & Yes \tabularnewline
 \hline
 3 & $\frac{1}{2}$ & 6.5 & $\frac{1}{96}(93+\sqrt{249})$ & $0.81$ &  $0.63$  & $\frac{1}{96}(81-\sqrt{561})$ & Yes \tabularnewline
   &               &     & $\frac{1}{96}(93+\sqrt{249})$ & $0.82$ &  $0.68$  & $\frac{1}{96}(83-\sqrt{409})$ & Yes \tabularnewline
 \hline
 3 & $\frac{1}{2}$ & 6.5 & $\frac{1}{96}(93-\sqrt{249})$ & $0.48$ &  $0.09$  & $0$           & No \tabularnewline
   &               &     & $\frac{1}{96}(93-\sqrt{249})$ & $0.57$ &  $0.31$  & $\frac{1}{4}$ & No \tabularnewline
 \hline
 \hline
 4 & $0$           & 6   & $\frac{5}{4}$ & $1.08$ &  $1.01$  & $1$ & Yes \tabularnewline
   &               &     & $\frac{5}{4}$ & $1.06$ &  $1.01$  & $1$ & Yes \tabularnewline
 \hline
 4 & $0$           & 6   & $\frac{5}{8}$ & $0.45$ &  $0.29$  & $\frac{1}{4}$ & No \tabularnewline
   &               &     & $\frac{5}{8}$ & $0.49$ &  $0.28$  & $\frac{1}{4}$ & No \tabularnewline
 \hline
 4 & $0$           & 6   & $\frac{1}{4}(3+\sqrt{2})$ & $0.74$ &  $0.54$  & $\frac{1}{2}$ & No  \tabularnewline
   &               &     & $\frac{1}{4}(3+\sqrt{2})$ & $0.81$ &  $0.66$  & $\frac{5}{8}$ & Yes \tabularnewline
 \hline
 4 & $0$           & 6   & $\frac{1}{4}(3-\sqrt{2})$ & $0.26$ &  $0.06$  & $0$ & No \tabularnewline
   &               &     & $\frac{1}{4}(3-\sqrt{2})$ & $0.30$ &  $0.08$  & $0$ & No \tabularnewline
 \hline
 \end{tabular}
 }
 \par \end{centering}
 \caption{Examples of evolution of $\mathbb{Z}_2$ orbifold pairs for different values of the parameter $\kappa = \frac{\check g}{g}$.
 }
\label{Pairskappa}
 \end{table}
 
More insight is obtained if we also look at the $\mathbb{Z}_2$ orbifold $(\check g = g)$. For the the orbifold theory (and for the whole interpolating theory with general $\check g$, $g$) we have an $SU(2)_L$ symmetry not present in $\Nm=2$ SCQCD, so to make the analysis more transparent we restrict the diagonalization to $SU(2)_L$ singlets. Our results for the $\mathbb{Z}_2$ orbifold are shown in Table \ref{Z2}. As in the case with maximal $r$-charge, parity pairs show up consistently. This is again expected because this theory is known to be integrable \cite{Beisert:2005he}.

Finally we look at how some sample parity pairs of the orbifold theory evolve when we
move away from the orbifold point.  Our results are shown in Table \ref{Pairskappa}. We see that for arbitrary values of $\kappa \equiv \check g/g$ the pairs are lifted and they are not in general recovered in the SCQCD limit $\kappa \to 0$.
(Note that not all $SU(2)_L$ gauge singlets evolve to legitimate states of ${\cal N}=2$ SCQCD, which must obey the stronger
 condition of being $SU(N_f)$ singlets. In the last column of Table \ref{Pairskappa} we indicate whether the states belong or not to ${\cal N}=2$ SCQCD.)

\subsection{$\Nm=1$ SQCD}

We now repeat the same analysis for $\Nm=1$ SQCD.
Inspired by the transformation used in the $\Nm=2$ theory, we define the following parity operation.
For the fields in the vector multiplet,
\begin{equation}
\begin{split}
A^a_{\mu b}  \lra -A^b_{\mu a}  \quad  \l^a_{\ph{a}b}  \lra -\l^b_{\ph{b} a}\, ,
\end{split}
\end{equation}
and analogous expressions for the conjugate fields.
For the chiral multipets, 
\begin{equation}
\begin{split}
\quad Q^{a i} \lra -\tilde Q_{\tilde \imath a}, \quad \psi^{a i} \lra -\tilde \psi_{\tilde \imath a}\, ,
\end{split}
\end{equation}
and analogous expressions for the conjugate fields. Again, these transformations are a symmetry of the Lagrangian and therefore commute with the dilation operator to all loops.
The action on single-trace states is  given by
\be 
P | A_1\, .\, .\, .\, A_L \rangle = (-1)^{L+k(k+1)/2}| A_L\, .\, .\, .\, A_1 \rangle\, ,
\ee
where $k$ is the number of fermions and we make the replacements $\psi  \lra \tilde \psi$, $\bar \psi  \lra \bar{\tilde{\psi}}$, $Q  \lra \tilde Q$, $\bar Q  \lra \bar{\tilde{Q}}$. Our results for the diagonalization of generalized single-trace operators of length $L \leq 5$ are shown in Table \ref{N1}.
We restrict to states with $r$-charge  $0 < r < L$. (We omit the sectors with $r=L$ and $r=0$, which are spanned respectively by $\{ \lambda_k \}$
and  $\{ \bar {\cal F}_k \}$. These sectors are isomorphic to the analogous sectors in ${\cal N}=4$ SYM and thus inherit their integrability.)

 \begin{table}[h]
 \begin{centering}
 {\footnotesize
 \begin{tabular}{|c|c|c|c|c|}
 \hline 
  $L$  & $r$  & $\Delta_0$  & $\delta \Delta^P\, [g_{YM}^2N/\pi^2]$ \tabularnewline
 \hline
  \hline 
 3  &  $1$             &  4.5  & $\frac{3}{4}^-$, $\frac{3}{8}^+$ \tabularnewline
    &                  &  5.5  & $\frac{1}{96}(67\pm\sqrt{457})^-$, $P_3(x)^+$\tabularnewline
    &                  &  6.5  & $\frac{25}{24}^-$, $\frac{1}{96}(81\pm\sqrt{473})^-$, $P_3(x)^+$ \tabularnewline
    &                  &       & $\frac{25}{48}^\pm$ \tabularnewline
    &  $2$             &  4    & $\frac{3}{16}^-$, $\frac{9}{16}^+$ \tabularnewline
    &                  &  5    & $\frac{7}{16}^-$, $\frac{7}{24}^-$, $\frac{13}{16}^+$, $\frac{1}{24}(13\pm\sqrt{39})^+$\tabularnewline
 \hline
 \hline 
 4  &  $1$             &  6.5  & $\frac{3}{4}^+$, $\frac{9}{8}^-$ \tabularnewline
    &                  &  7.5  & $P_3(x)^-$, $P_4(x)^+$ \tabularnewline
    &  $2$             &  6    & $\frac{1}{16}(9\pm\sqrt{37})^+$ , $\frac{1}{96}(67\pm\sqrt{457})^+$ , $P_3(x)^-$ \tabularnewline
    &                  &       & $\frac{9}{16}^\pm$ , $0^\pm$  \tabularnewline
    &                  &   7   & $\frac{1}{96}(81\pm\sqrt{473})^+$, $P_3(x)^-$, $P_7(x)^+$, $P_8(x)^-$ \tabularnewline
    &                  &       & $\frac{25}{48}^\pm$\tabularnewline
    &  $3$             &  5.5  & $\frac{7}{16}^-$, $\frac{13}{16}^+$ \tabularnewline
    &                  &  6.5  & $\frac{11}{16}^+$, $\frac{17}{16}^-$, $\frac{1}{96}(71\pm\sqrt{553})^+$, $P_3(x)^-$   \tabularnewline
 \hline
 \hline 
 5  &  $3$   &  6.5  & $\frac{1}{16}(9\pm\sqrt{37})^-$, $\frac{9}{16}^\pm$   \tabularnewline
 \hline
 \hline
   \multicolumn{4}{|c|}{Paired eigenvalues: $\sim$ 13 \%} \tabularnewline
 \hline
 \end{tabular}
 }
 \par \end{centering}
 \caption{ $SU(1,1)$ primaries with  $0 < r < L$
  in the $SU(1,1) \times U(1|1)$ sector of ${\cal N}=1$ SQCD.}
\label{N1}
 \end{table}
The results are qualitatively
similar to the ones for ${\cal N}=2$ SCQCD: there are a few parity pairs, but their presence is not as striking and systematic
as in ${\cal N}=4$ SYM.

\section{Discussion}
\label{discussion}

We have used superconformal invariance to constrain the planar one-loop dilation operator
 for ${\cal N}=1$ SQCD.
The structure of the calculation is similar to the one in ${\cal N}=2$ SCQCD \cite{Liendo:2011xb}, and leads
to the same surprising result: the one-loop Hamiltonian is uniquely fixed by symmetry.
We have worked in the ``electric'' description of the theory, at the Banks-Zaks fixed point near the
upper edge of the conformal window. It would be interesting to apply the same
strategy to the dual magnetic theory, at the Banks-Zaks fixed point
in the lower edge of the conformal window.
 
The  recent discovery   \cite{Poland:2011kg}  that  in the scalar sector of  ${\cal N}=1$ SQCD
 the planar one-loop dilation operator is captured by
 the integrable Ising chain in transverse magnetic field
was one of the motivations of our work. 
  A skeptic may  point out  that this identification
  was  kinematically foreordained, since no other structure
  is allowed (though the precise value of the transverse magnetic field does require dynamical input). The same could be said
   about the $SU(2)$ scalar sector of ${\cal N}=4$ SYM, which is {\it a priori} guaranteed to take the form of the Heisenberg spin chain, but not (say) for the $SO(6)$ sector
considered in the seminal paper \cite{Minahan:2002ve}. 
More convincing tests of one-loop integrability require the complete Hamiltonian, which we have obtained in this paper.
Our analysis indicates that the presence of degenerate parity pairs, a hallmark of integrability,
is not as systematic  in 
${\cal N}=1$ SQCD (or in ${\cal N}=2$ SCQCD) as it is
  in ${\cal N}=4$ SYM. These preliminary results are  not particularly encouraging, but 
 should of course be taken {\it cum grano salis}.

 If the complete theory  turns out to be non-integrable, we can still ask whether there
 is scope for integrability in some closed subsectors. The question is really about
 all-loop integrability, since at one loop some simple sectors may be ``accidentally'' integrable.
 This may well be the case for the scalar sector of \cite{Poland:2011kg}, which is only closed
 to lowest order: at higher orders the scalars mix with every other state,
 so either the whole theory turns out to be integrable or the integrability of the scalar spin chain is a one-loop accident.
More promising is the situation in the
 sectors that happen to coincide at one-loop with analogous sectors
 of ${\cal N}=4$ SYM,  thus inheriting its integrability properties,
 but that remain closed to all loops.
 The largest such sector is the $SU(2,1|1)$ sector spanned by the letters $\{ {\cal D}_{+ \dot \alpha}^k\, \lambda_{+}\,,\; {\cal D}_{+ \dot \alpha}^k \,\Fm_{++} \}$.
 While at one loop its Hamiltonian coincides  with that of ${\cal N}=4$ SYM, it will start differing from it at sufficiently high order. It will be very interesting to investigate whether integrability is preserved.\footnote{The same  can be asked 
 for the $SU(2,1|2)$ sector of the ${\cal N}=2$ interpolating theory (in particular of ${\cal N}=2$ SCQCD)
consisting of the letters  $\{{\cal D}_{+ \dot \alpha}^k\, \phi\, , {\cal D}_{+ \dot \alpha}^k\, \lambda_{\Im \,+}\,,\; {\cal D}_{+ \dot \alpha}^k \,\Fm_{++} \}$.
Here the question is even sharper since on the $\phi$ vacuum the two-body magnon S-matrix is completely fixed by symmetry (up to the overall dynamical phase)
to be the $SU(2|2)$ S-matrix of \cite{Beisert:2005tm}, which automatically satisfies the Yang-Baxter equation. Still, integrability is by no means obvious, since factorization of the $n$-body S-matrix is a stronger condition than Yang-Baxter. If this sector turns out to be integrable to all loops, its difference from the analogous sector
of  ${\cal N}=4$ SYM would  be fully encoded in the expression
of the dynamical phase and of the magnon dispersion relation. The difference with ${\cal N}=4$ SYM should start appearing at three loops \cite{Pomoni:2011}.} The $SU(2,1|1)$ sector exists
of course also in the dual magnetic theory, so its integrability
may allow to interpolate across  the whole conformal window. We  look forward to future investigations of this scenario.

\section*{Acknowledgements}
It is a pleasure to thank  David Poland for useful conversations.
The  work of P.L. and L.R. was supported in part  by NSF grant  PHY-1066293.
Any opinions, findings, and conclusions or recommendations expressed in this material are those of the authors and do not necessarily reflect the views of the National Science Foundation.

\newpage

\appendix
\section{$\Nm=1$ superconformal multiplets}

In this appendix we summarize some basic facts about $\Nm=1$
superconformal representation theory \cite{Dolan:2008qi}. 
 A generic long multiplet
$\AA^{\Delta}_{r(j_1, j_2)}$ is generated by the action of $4$
Poincar\'e supercharges $\QQ_\alpha$ and ${\bar{\Qm}}_{\dot \alpha}$
on a superconformal primary which by definition is annihilated by
all the conformal supercharges $\cal S$.
In Table \ref{N1-shortening} we have
summarized the possible shortening and semi-shortening conditions.

\begin{center}
\begin{table}[!h]
{\small
\begin{centering}
\begin{tabular}{|l|l|l|l|l|}
\hline
\multicolumn{4}{|c|}{Shortening Conditions} & Multiplet\tabularnewline
\hline
\hline
$\BB$ & $\QQ_{\alpha}|r\rangle^{h.w.}=0$ & $j_1=0$ & $\Delta=-\frac{3}{2}r$ & $\BB_{r(0,j_2)}$\tabularnewline
\hline
$\bar{\BB}$ & $\bar{\QQ}_{\dot{\alpha}}|r\rangle^{h.w.}=0$ & $j_2=0$ & $\Delta=\frac{3}{2}r$ & $\bar{\BB}_{r(j_1,0)}$\tabularnewline
\hline
$\hat{\BB}$ & $\BB\cap\bar{\BB}$ & $j_1,j_2,r=0$ & $\Delta=0$ & $\hat{\BB}$\tabularnewline
\hline
\hline
$\CC$ & $\e^{\alpha\beta}\QQ_{\beta}|r\rangle_{\alpha}^{h.w.}=0$ &  & $\Delta=2+2j_1-\frac{3}{2}r$ & $\CC_{r(j_1,j_2)}$\tabularnewline
 & $(\QQ)^{2}|r\rangle^{h.w.}=0$ for $j_1=0$ &  & $\Delta=2-\frac{3}{2}r$ & $\CC_{r(0,j_2)}$\tabularnewline
\hline
$\bar{\CC}$ & $\e^{\dot{\alpha}\dot{\beta}}\bar{\QQ}_{\dot{\beta}}|r\rangle_{\dot{\alpha}}^{h.w.}=0$ &  & $\Delta=2+2j_2+\frac{3}{2}r$ & $\bar{\CC}_{r(j_1,j_2)}$\tabularnewline
 & $(\bar{\QQ})^{2}|r\rangle^{h.w.}=0$ for $j_2=0$ &  & $\Delta=2+\frac{3}{2}r$ & $\bar{\CC}_{r(j_1,0)}$\tabularnewline
\hline
$\hat{\CC}$ & $\CC\cap\bar{\CC}$ & $\frac{3}{2}r=(j_1-j_2)$ & $\Delta=2+j_1+j_2$ & $\hat{\CC}_{(j_1,j_2)}$\tabularnewline
\hline
$\DD$ & $\BB\cap\bar{\CC}$ & $j_1=0,-\frac{3}{2}r=j_2+1$ & $\Delta=-\frac{3}{2}r=1+j_2$ & $\DD_{(0,j_2)}$\tabularnewline
\hline
$\bar{\DD}$ & $\bar{\BB}\cap\CC$ & $j_2=0,\frac{3}{2}r=j_1+1$ & $\Delta=\frac{3}{2}r=1+j_1$ & $\bar{\DD}_{(j_1,0)}$\tabularnewline
\hline
\end{tabular}
\par\end{centering}
} \caption{\label{N1-shortening}Possible shortening conditions 
for the $\NN=1$ superconformal algebra.}
\end{table}
\par\end{center}

\section{Oscillator Representation}
\label{oscillatorsA}

In this appendix we describe the oscillator representation of the $\Nm=1$ superconformal algebra $SU(2,2|1)$.
We introduce two sets of bosonic oscillators $(\ab^\a,\ab^\dag_\a)$, $(\bb^{\ad},\bb^\dag_{\ad})$ and one
fermionic oscillator
$(\cb,\cb^\dag)$, where $(\a,\ad)$ are Lorentz indices. In addition  we will 
need three more ``auxiliary'' fermionic operators $(\db_i,\db^{\dag}_i)$, $i=1,2,3$.
The non-zero (anti)commutation relations are
\bea 
[\ab^\a,\ab^\dag_\b] & = & \delta^\a_\b\, ,
\\
\lbr \bb^{\ad},\bb^\dag_{\bd} \rbr & = & \delta^{\ad}_{\bd}\, ,
\\
\{ \cb,\cb^\dag\} & = & 1\, ,
\\ 
\{ \db_i,\db^\dag_j \} & = & \delta_{ij}\, .
\eea
In this oscillator representation the generators of $SU(2,2|1)$ are given by

  \begin{align}
  & &\Lm^{\ph{\b}\a}_{\b} & = \ab^\dag_\b\ab^\a-\frac{1}{2}\delta^\a_\b \ab^\dag_\g\ab^\g\, , 
  \\
  & &\dot{\Lm}^{\ph{\bd}\ad}_{\bd} & =  \bb^\dag_{\bd}\bb^{\ad}-\frac{1}{2}\delta^{\ad}_{\bd} \bb^\dag_{\gd}\bb^{\gd}\, , 
  \\
  & & r & =   \cb^\dag\cb-\frac{1}{3}\db_1^\dag\db_1-\frac{1}{3}\db_2^\dag\db_2-\db_3^\dag\db_3\, , 
  \\
  & & D & =  1 + \frac{1}{2}\ab_{\g}^{\dag}\ab^\g + \frac{1}{2}\bb_{\gd}^{\dag}\bb^{\gd}\, , 
  \\
  & & C & =  1 - \frac{1}{2}\ab_{\g}^{\dag}\ab^\g + \frac{1}{2}\bb_{\gd}^{\dag}\bb^{\gd}-\frac{1}{2}\cb^\dag\cb
 -\frac{1}{2}\db_1^\dag\db_1 -\frac{1}{2}\db_2^\dag\db_2 -\frac{3}{2}\db_3^\dag\db_3\,, &
 \end{align}
 \begin{align}
   \Qm_{\a}   & = \ab^{\dag}_{\a}\cb\, , &  \bar{\Qm}_{\ad} & =  \bb^{\dag}_{\ad}\cb^{\dag}\, , 
  \\
   \Sm^{\a}   & = \cb^{\dag}\ab^{\a}\, , & \bar{\Sm}^{\ad} & = \bb^{\ad}\cb\, , 
  \\  
    \Pm_{\a \bd}          & = \ab^{\dag}_{\a} \bb^{\dag}_{\bd}\, , &  \Km^{\a \bd} & = \ab^\a \bb^{\bd}\, ,
  \end{align}
 Here $C$ is a central charge that must kill any physical state. It could be eliminated from the algebra by redefining $r\rightarrow r+\frac{2}{3}C$, but it is useful for implementing the harmonic action so we keep it.

\subsection{Vector multiplets $\Vm$ and $\bar{\Vm}$}

We define a vacuum state $|0\rangle$ annihilated by all the lowering operators. Then we identify 
\bea 
\Dm^k \Fm & = & (\ab^{\dag})^{k+2} (\bb^{\dag})^k  |0\rangle\, , \\
\Dm^k\lambda & = & (\ab^{\dag})^{k+1} (\bb^{\dag})^k \cb^{\dag}|0\rangle\, ,
\eea
and
\bea 
\Dm^k \bar{\Fm} & = & (\ab^{\dag})^{k} (\bb^{\dag})^{k+2} \cb^{\dag} \db_3^{\dag} |0\rangle\, ,\\
\Dm^k\bar{\lambda} & = & (\ab^{\dag})^{k} (\bb^{\dag})^{k+1} \db_3^{\dag} |0\rangle\, .
\eea

\subsection{Chiral multiplets $\Xm$ and $\bar{\Xm}$}

Similarly, for the chiral multiplets we identify
\bea 
\Dm^k Q & = & (\ab^{\dag})^{k} (\bb^{\dag})^k \cb^{\dag} \db_1^{\dag}|0\rangle\, ,
\\
\Dm^k \psi & = & (\ab^{\dag})^{k+1} (\bb^{\dag})^k \db_1^{\dag} |0\rangle\, ,
\eea
and
\bea
\Dm^k \bar{Q} & = & (\ab^{\dag})^{k} (\bb^{\dag})^k \db_1^{\dag}\db_2^{\dag}|0\rangle\, ,
\\
\Dm^k \bar{\psi} & = & (\ab^{\dag})^{k} (\bb^{\dag})^{k+1} \cb^{\dag}\db_1^{\dag}\db_2^{\dag}|0\rangle\, .
\eea

\subsection{Two-letter Superconformal Primaries}
\label{PrimariesA}

By demanding that they are annihilated by all the conformal supercharges and by the
appropriate combinations of Poincar\'e supercharges, 
we have worked out the expressions for the superconformal primaries of the
irreducible modules that appear on the right hand side of  the tensor products (\ref{XtXT}--\ref{VVbT}).
The grassmannOps.m oscillator package by Jeremy Michelson and Matthew Headrick was extremely useful for this task.
We simply quote the results:

\underline{\textbf{$\Vm \times \Vm$:}}
\bea
\bar{\Bm}_{2(0,0)} & = & \l_+ \l_--\l_- \l_+\, ,
\\
\label{VVprimaryq0}
\bar{\Bm}_{2(1,0)} & = & \l_+ \l_+\, ,
\\
\nn
\hat{\Cm}_{(\frac{q+1}{2},\frac{q-2}{2})} & = & \sum_{k=0}^{q-2} \frac{(-1)^k}{q(q+1)}\binom{q+1}{k+2}\binom{q-2}{k}
\Dm^{q-k-2}\lambda_{+} \Dm^k\Fm_{++}
\\
\label{VVprimaryq}
& & + \frac{1}{q}\sum_{k=0}^{q-2} \frac{(-1)^{q-k}}{k+2}\binom{q-2}{k}\binom{q}{k+1}
\Dm^{k}\Fm_{++} \Dm^{q-k-2}\l_+\, .
\eea
For $\bar{\Vm}\times \bar{\Vm}$ the expressions are identical with $(\l,\Fm)$ replaced by $(\bar{\l},\bar{\Fm})$.

\underline{\textbf{$\Vm \times \Xm$:}}
\bea
\label{VXprimaryqm1}
\bar{\Bm}_{\frac{5}{3}(\frac{1}{2},0)} & = & \l_+ Q \, ,
\\
\nn
\hat{\Cm}_{(\frac{q+1}{2},\frac{q-1}{2})} & = & \sum_{k=0}^{q-1} (-1)^k\binom{q-1}{k}\binom{q+1}{k}
\Dm^{q-k-1}\Fm_{++} \Dm^{k}Q
\\
\label{VXprimaryq}
&  & +(q+1)\sum_{k=0}^{q-1}\frac{(-1)^{k}}{k+1}\binom{q-1}{k}\binom{q}{k} \Dm^{q-k-1}\l_{+}\Dm^{k}\psi_+\,.
\eea
For the $\tilde{\Xm} \times \Vm$ primary we replace ($Q$,$\psi$) by ($\tilde{Q}$,$\tilde{\psi}$) and interchange the order of the fields (taking into account fermionic minus signs).

\underline{\textbf{$\bar{\Vm} \times \Xm$:}}
\bea
\hat{\Cm}_{(\frac{q}{2},\frac{q+1}{2})} & = & \sum_{k=0}^{q} (-1)^k\binom{q}{k}\binom{q+1}{k}
\Dm^{q-k}\bar{\l}_{\dot{+}} \Dm^{k}Q
\\
\label{VbXprimaryq}
&  & -q\sum_{k=0}^{q-1}\frac{(-1)^{k}}{k+1}\binom{q-1}{k}\binom{q+1}{k} \Dm^{q-k-1}\bar{\Fm}_{\dot{+}\dot{+}}\Dm^{k}\psi_+\,.
\eea
For the $\tilde{\Xm} \times \bar{\Vm}$ primary we replace ($Q$,$\psi$) by ($\tilde{Q}$,$\tilde{\psi}$) and interchange the order of the fields.

\underline{\textbf{$\bar{\Xm} \times {\Vm}$:}}
\bea
\nn
\hat{\Cm}_{(\frac{q+1}{2},\frac{q}{2})} & = & \sum_{k=0}^{q} (-1)^k\binom{q}{k}\binom{q+1}{k+1}
\Dm^{q-k}\bar{Q} \Dm^{k}\l_+
\\
\label{XbVprimaryq}
&  & +q\sum_{k=0}^{q-1} \frac{(-1)^{k}}{k+2}\binom{q-1}{k}\binom{q+1}{k+1} \Dm^{q-k-1}\bar{\psi}_{\dot{+}} \Dm^{k}\Fm_{++}\, .
\eea
For the $\Vm \times \bar{\tilde{\Xm}}$ primary we replace ($Q$,$\psi$) by ($\tilde{Q}$,$\tilde{\psi}$) and interchange the order of the fields.

\underline{\textbf{$\bar{\Xm} \times \bar{\Vm}$:}}
\bea
\label{XbVbprimaryqm1}
\Bm_{-\frac{5}{3}(0,\frac{1}{2})} & = & \bar{Q}\bar{\l}_+ \, ,
\\
\nn
\hat{\Cm}_{(\frac{q-1}{2},\frac{q+1}{2})}  & = & \sum_{k=0}^{q-1}\frac{(-1)^k}{k+1}\binom{q-1}{k}\binom{q}{k}
\Dm^{k}\bar{\psi}_{\dot{+}} \Dm^{q-k-1}\bar{\l}_{\dot{+}}
\\
\label{XbVbprimaryq0}
&  & +\sum_{k=0}^{q-1}\frac{(-1)^{q-k}}{q+1}\binom{q-1}{k}\binom{q+1}{k+2}
\Dm^{q-k-1}\bar{Q} \Dm^{k}\bar{\Fm}_{\dot{+}\dot{+}}\,.
\eea
For the $\bar{\Vm} \times \bar{\tilde{\Xm}}$ primary we replace ($Q$,$\psi$) by ($\tilde{Q}$,$\tilde{\psi}$) and interchange the order of the fields.

\underline{\textbf{$\bar{\Xm} \times \Xm$:}}
\bea
\nn
\hat{\Cm}_{(\frac{q}{2},\frac{q}{2})}  & = & \sum_{k=0}^{q}(-1)^k\binom{q}{k}\binom{q}{k}
\Dm^{q-k}\bar{Q} \Dm^{k}Q
\\
\label{XbXprimaryq}
&  & +q\sum_{k=0}^{q-1}\frac{(-1)^{k}}{k+1}\binom{q-1}{k}\binom{q}{k}
\Dm^{q-k-1}\bar{\psi}_{\dot{+}} \Dm^{k}\psi_+\,.
\eea
This primary is \textit{gauge} contracted. For the \textit{flavor} contracted
$\bar{\tilde{\Xm}} \times \tilde{\Xm}$ primary we replace ($Q$,$\psi$) by ($\tilde{Q}$,$\tilde{\psi}$).

\underline{\textbf{$\tilde{\Xm} \times \bar{\tilde{\Xm}}$:}}
\bea
\nn
\hat{\Cm}_{(\frac{q}{2},\frac{q}{2})}  & = & \sum_{k=0}^{q}(-1)^k\binom{q}{k}\binom{q}{k}
\Dm^{q-k}\tilde{Q} \Dm^{k}\bar{\tilde{Q}}
\\
\label{XtXtbprimaryq}
&  & +q\sum_{k=0}^{q-1}\frac{(-1)^{k}}{k+1}\binom{q-1}{k}\binom{q}{k}
\Dm^{q-k-1}\tilde{\psi}_{+} \Dm^{k}\bar{\tilde{\psi}}_{\dot{+}}\,.
\eea
This primary is \textit{gauge} contracted. For the \textit{flavor} contracted
$\Xm \times \bar{\Xm}$ primary we replace ($\tilde{Q}$,$\tilde{\psi}$) by ($Q$,$\psi$) .

\underline{\textbf{$\tilde{\Xm} \times \Xm$:}}
\bea
\bar{\Bm}_{\frac{4}{3}(0,0)}  & = & \tilde{Q}Q
\\
\nn
\hat{\Cm}_{(\frac{q+1}{2},\frac{q}{2})}  & = & \sum_{k=0}^{q}(-1)^k\binom{q}{k}\binom{q+1}{k+1}
\Dm^{q-k}\tilde{Q} \Dm^{k}\psi_+
\\
\label{XtXprimaryq}
&  & -\sum_{k=0}^{q}(-1)^{q-k}\binom{q}{k}\binom{q+1}{k+1}
\Dm^{k}\tilde{\psi}_{+} \Dm^{q-k}Q\,.
\eea
For the $\bar{\Xm} \times \bar{\tilde{\Xm}}$ primary we interchange ($\tilde{Q}$,$\psi$) by ($\bar{Q}$,$\bar{\tilde{\psi}}$) (also for the conjugates).

\underline{\textbf{$\Vm \times \bar{\Vm}$:}}
\bea
\nn
\hat{\Cm}_{(\frac{q}{2},\frac{q}{2})} & = & \sum_{k=0}^{q-1}\frac{(-1)^k}{k+1}\binom{q-1}{k}\binom{q}{k}
 \Dm^{q-k-1}\l_+ \Dm^{k}\bar{\l}_{\dot{+}}
\\
\label{VVbprimaryq}
&  & -\sum_{k=0}^{q-2}\frac{(-1)^{q-k}}{q}\binom{q1}{k}\binom{q}{k+2} \Dm^{k}\Fm_{++} \Dm^{q-k-2}\bar{\Fm}_{\dot{+}\dot{+}}\,.
\eea
For the $\bar{\Vm} \times \Vm$ primary we replace ($\l$,$\Fm$) by($\bar{\l}$,$\bar{\Fm}$) .

%\newpage

\bibliographystyle{JHEP}
\bibliography{Completebbl}

\end{document}